\documentclass[twocolumn]{aastex63}
\usepackage{multirow}
\usepackage[T1]{fontenc}
\usepackage{mhchem}
\usepackage{url}
\newcommand{\Htwo}{\(\textrm{H}_{2}\)}
\newcommand{\CO}{\(^{12}\textrm{CO}\)}
\newcommand{\thCO}{\(^{13}\textrm{CO}\)}
\newcommand{\CetO}{\(\textrm{C}^{18}\textrm{O}\)}
\newcommand{\HtwoO}{\(\textrm{H}_{2}\textrm{O}\)}
\newcommand{\Msun}{\(\textrm{M}_{\odot}\)}

\received{some date, 2019}
\revised{some date, 2019}
\accepted{some day}
\submitjournal{AJ}

\shorttitle{TW Hya Thermal Profile and Disk Mass}

\begin{document}

\title{The TW Hya Rosetta Stone Project III: Resolving the Gaseous Thermal Profile of the Disk}

\author[0000-0002-0150-0125]{Jenny K. Calahan}
\affiliation{University of Michigan, 323 West Hall, 1085 South University Avenue, Ann Arbor, MI 48109, USA }

\author[0000-0003-4179-6394]{Edwin Bergin}
\affiliation{University of Michigan, 323 West Hall, 1085 South University Avenue, Ann Arbor, MI 48109, USA }

\author[0000-0002-0661-7517]{Ke Zhang}
\altaffiliation{NASA Hubble Fellow}
\affiliation{Department of Astronomy, University of Wisconsin-Madison, 
475 N Charter St, Madison, WI 53706}
\affiliation{University of Michigan, 323 West Hall, 1085 South University Avenue, Ann Arbor, MI 48109, USA }

\author[0000-0003-1534-5186]{Richard Teague}
\affiliation{Center for Astrophysics | Harvard \& Smithsonian, 60 Garden Street, Cambridge, MA 02138, USA}

\author[0000-0003-2076-8001]{Ilsedore Cleeves}
\affiliation{Astronomy Department, University of Virginia, Charlottesville, VA 22904, USA}

\author[0000-0002-8716-0482]{Jennifer Bergner}
\affiliation{University of Chicago, Department of the Geophysical Sciences, Chicago, IL 60637, USA}

\author[0000-0003-0787-1610]{Geoffrey A. Blake}
\affiliation{Division of Chemistry \& Chemical Engineering, California Institute of Technology, Pasadena CA 91125}

\author[0000-0002-1917-7370]{Paolo Cazzoletti}
\affiliation{Leiden Observatory, Leiden University, 2300 RA Leiden, The Netherlands}

\author[0000-0003-4784-3040]{Viviana Guzm\'an}
\affiliation{Instituto de Astrofísica, Ponticia Universidad Católica de Chile, Av. Vicuña Mackenna 4860, 7820436 Macul, Santiago, Chile}

\author[0000-0001-5217-537X]{Michiel R.Hogerheijde}
\affiliation{Leiden Observatory, Leiden University, 2300 RA Leiden, The Netherlands}
\affiliation{Anton Pannekoek Institute for Astronomy, University of Amsterdam, The Netherlands.}

\author[0000-0001-6947-6072]{Jane Huang}
\altaffiliation{NHFP Sagan Fellow}
\affiliation{Center for Astrophysics | Harvard \& Smithsonian, 60 Garden Street, Cambridge, MA 02138, USA}
\affiliation{University of Michigan, 323 West Hall, 1085 South University Avenue, Ann Arbor, MI 48109, USA }

\author[0000-0003-0065-7267]{Mihkel Kama}
\affiliation{Tartu Observatory, University of Tartu, Observatooriumi 1, 61602, T\~{o}ravere, Estonia}
\affiliation{Institute of Astronomy, University of Cambridge, Madingley Road, Cambridge CB3 0HA, UK}

\author[0000-0002-8932-1219]{Ryan Loomis}
\affiliation{National Radio Astronomy Observatory (NRAO), 520 Edgemont Rd, Charlottesville, VA 22903, USA}

\author[0000-0001-8798-1347]{Karin {\"O}berg}
\affiliation{Center for Astrophysics | Harvard \& Smithsonian, 60 Garden Street, Cambridge, MA 02138, USA}

\author[0000-0001-7591-1907]{Ewine F. van Dishoeck}
\affiliation{Max-Planck-Institut für Extraterrestrische Physik, Giessenbachstrasse1, 85748 Garching, Germany}
\affiliation{Leiden Observatory, Leiden University, 2300 RA Leiden, The Netherlands}

\author[0000-0002-3800-9639]{Jeroen Terwisscha van Scheltinga}
\affiliation{Leiden Observatory, Leiden University, PO Box 9513, 2300 RA Leiden, The Netherlands}
\affiliation{Laboratory for Astrophysics, Leiden Observatory, Leiden University, PO Box 9513, 2300 RA Leiden, The Netherlands}

\author[0000-0001-6078-786X]{Catherine Walsh}
\affiliation{School of Physics and Astronomy
University of Leeds, Leeds LS2 9JT, UK}

\author[0000-0003-1526-7587]{David Wilner}
\affiliation{Center for Astrophysics | Harvard \& Smithsonian, 60 Garden Street, Cambridge, MA 02138, USA}

\author[0000-0001-8642-1786]{Charlie Qi}
\affiliation{Center for Astrophysics | Harvard \& Smithsonian, 60 Garden Street, Cambridge, MA 02138, USA}



\begin{abstract}

The thermal structure of protoplanetary disks is a fundamental characteristic of the system that has wide reaching effects on disk evolution and planet formation. In this study, we constrain the 2D thermal structure of the protoplanetary disk TW Hya structure utilizing images of seven CO lines. This includes new ALMA observations of \CO{} $J$=2-1 and \CetO{} $J$=2-1 as well as archival ALMA observations of \CO{} $J$=3-2, \thCO{} $J$=3-2, 6-5, \CetO{} $J$= 3-2, 6-5. Additionally, we reproduce a \textit{Herschel} observation of the HD $J$=1-0 line flux, the spectral energy distribution, and utilize a recent quantification of CO radial depletion in TW Hya. These observations were modeled using the thermo-chemical code RAC2D, and our best fit model reproduces all spatially resolved CO surface brightness profiles. The resulting thermal profile finds a disk mass of 0.025~\(\textrm{M}_{\odot}\) and a thin upper layer of gas depleted of small dust with a thickness of \(\sim\)1.2\% of the corresponding radius. Using our final thermal structure, we find that CO alone is not a viable mass tracer as its abundance is degenerate with the total \Htwo{} surface density.  Different mass models can readily match the spatially resolved CO line profiles with disparate abundance assumptions. Mass determination requires additional knowledge and, in this work, HD provides the additional constraint to derive the gas mass and support the inference of CO depletion in the TW Hya disk. Our final thermal structure confirms the use of HD as a powerful probe of protoplanetary disk mass. Additionally, the method laid out in this paper is an employable strategy for extraction of disk temperatures and masses in the future. 

\end{abstract}



\section{Introduction} \label{sec:intro}

The radial and vertical (2D) thermal profile of a gaseous protoplanetary disk is difficult to uncover but has wide reaching effects on the physics, chemistry, and thus the planet formation potential of a disk. Further, temperature is often essential for estimation of other fundamental disk properties, such as the local sound speed and disk mass \citep{Bergin17}. Any physical process that relies on sound speed will also be affected by temperature. Turbulent viscosity, as one example, relies on sound speed and plays an important role in the transportation and redistribution of disk material \citep{Shakura1973}. The vertical density structure, in addition to the radial dependence, gives rise to the flaring of the disk, and is highly dependent on the thermal structure  \citep{Kenyon95}. In turn, the level of flaring sets the angle of incidence to stellar irradiation, producing strong vertical thermal gradients that lead to density profiles that deviate significantly from the derived Gaussian density profiles from assuming a vertically isothermal disk \citep{Aikawa02,Gorti11,Woitke09}.

Temperature is also a determinate parameter to chemical processes. The gas temperature influences the rate of gas-phase exothermic reactions  and, in particular, reactions that possess a significant activation barrier. Further, the midplane temperature controls the balance between gas-phase deposition and sublimation and thus  the relative spatial composition of ices. For example, \HtwoO{} freezes out at dust temperatures lower than $\approx$ 120 K - 170 K \citep{Fraser01,Bergin18}, while CO freezes out below \(\sim\)21-25 K in TW Hya, \citep{Bisschop06,Fayolle16,Schwarz16, Zhang17}. 
Thus the gas/ice transition or snowline for water and CO is set by the thermal structure with water close to the star and CO at greater distances.
Temperature-dependent snowlines are theorized to potentially be favorable sites for  planet formation \citep{Hayashi81,Stevenson88, Zhang15, schoonenberg17}.  Further, across various snowlines of key elemental carriers, the relative chemical composition  of the disk changes the gas/solid state balance of these carriers, which will directly influence the chemical composition of planetary atmospheres or cores at birth \citep[e.g. C/O ratio,][]{Oberg11, Oberg16}. 

\begin{figure*}
    \includegraphics[width=\textwidth]{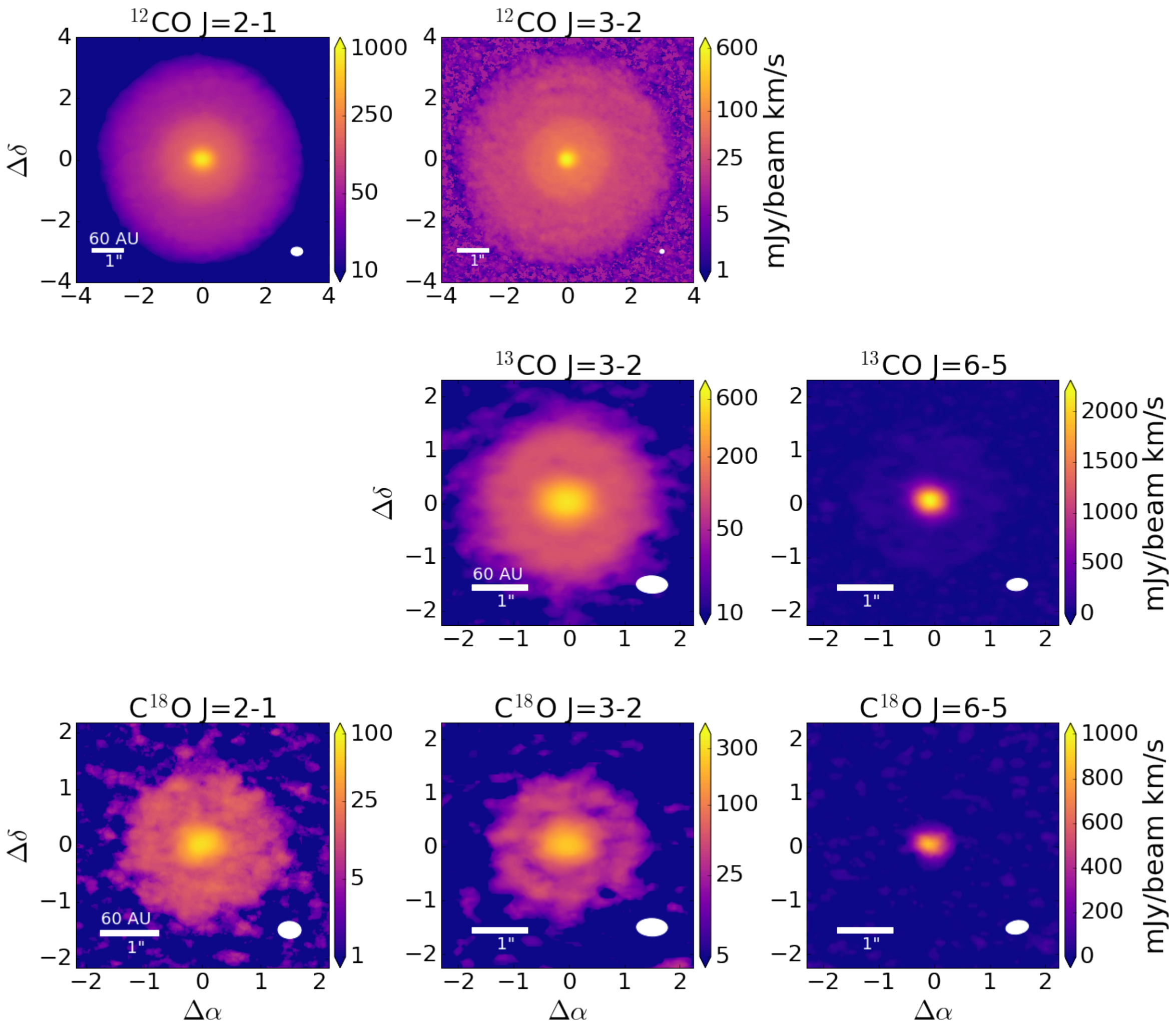}
    \caption{Integrated emission maps of TW Hya in \CO{} J=2-1, 3-2, \thCO{} J=3-2, 6-5, \CetO{} J=2-1, 3-2, and 6-5. The beam size is in the bottom right of each image. All lines of CO with the exception of the 6-5 lines are shown in log scale to amplify features. The 6-5 lines are shown on a linear scale as they have one peak and no substructure to highlight. The contour lines shown in \thCO{} 3-2 and \CetO{} 6-5 are overlain on \CO{} 2-1 to compare emission region sizes. }  
    \label{obs}
\end{figure*}
\renewcommand{\arraystretch}{1.2}
\begin{deluxetable*}{cccccccccc}
    \tablecolumns{7} 
    \tablewidth{0pt}
    \tablecaption{ALMA Observations Summary}
     \tablehead{&&&&&&\(\sigma ^{b}\) & v\(_\sigma^{c}\)& \(\rm I_{peak}^{d}\) & \(\rm F_{\textrm{line}}^{e}\)\\
     Program ID & P.I.  & Species and & Frequency & \(\rm E_{\textrm{up}}^{a}\)& Beam [PA] & (mJy/& (km/&(mJy/bm & (Jy \\
     &&Transition& (GHz) &(K)&(AU \(\times\) AU) [$^{\circ}$]& bm) & s)& km/s) & km/s)}
    \startdata
    2016.1.00311.S & I. Cleeves & \CO{} J=2-1 & 230.538& 16.6 & 22 \(\times\) 16 [89.02] & 3.9 & 0.56 & 828 & 17.8\\
    2015.1.00686.S & S. Andrews & \multirow{2}{*}{\CO{} J=3-2} &  \multirow{2}{*}{345.795} & \multirow{2}{*}{33.2} & \multirow{2}{*}{8.3 \(\times\) 7.7 [-74.96]} & \multirow{2}{*}{1.7} & \multirow{2}{*}{0.25} & \multirow{2}{*}{541}&\multirow{2}{*}{43.2}\\
    2016.1.00629.S & I. Cleeves&&&&&&&\\
    2012.1.00422.S & E. Bergin & \thCO{} J=3-2 & 330.587 & 15.9 & 30 \(\times\) 18 [88.15] & 9.2 & 0.09 & 574 & 4.35\\
    2012.1.00422.S & E. Bergin & \thCO{} J=6-5 & 661.067 & 111.1 & 23 \(\times\) 14 [-84.84]& 56 & 0.11 & 2248 & 8.21\\
    2016.1.00311.S & I. Cleeves & \CetO{} J=2-1 & 219.560 & 15.8 & 24 \(\times\) 18 [87.65]& 3.2 & 0.10 & 66 & 0.57\\
    2012.1.00422.S & E. Bergin & \CetO{} J=3-2 & 329.331 & 31.6	& 30 \(\times\) 18 [88.62]& 12 & 0.11 &252 &1.16\\
    2012.1.00422.S & E. Bergin & \CetO{} J=6-5 & 658.553 &110.6& 23 \(\times\) 14 [-79.98]& 77 & 0.09 & 973& 1.56\\
    \enddata
    \tablenotetext{a}{\citet{Muller05}}
    \tablenotetext{b}{RMS Noise -- measured from line free channels}
    \tablenotetext{c}{Velocity width assumed in RMS determination}
    \tablenotetext{d}{Peak Integrated Intensity Flux -- errors are visualized in Figure \ref{COprof}}
    \tablenotetext{e}{Total Integrated Flux, calculated within a circle with radius 4`` for \CO{} 2-1 and 3-2; circle with radius 2``  for \CetO{} 2-1, 3-2, and \thCO{} 3-2; and a circle with radius 1`` for \CetO{}, \thCO{} 6-5  (See Figure \ref{obs}). The error in the observed flux ranges from 10-15\% based on the ALMA technical handbook.}
    \label{ALMA}
\end{deluxetable*}

Temperature is also essential in estimating one of the most fundamental properties of a protoplanetary disk: its mass. Assuming a constant mass ratio between gas and dust and a dust mass opacity, sub-mm thermal continuum emission is generally used as an approximate tracer for the gas mass \citep{Bergin17}. However, this estimation is subject to large uncertainty as grain evolution alters the dust opacity, both spatially and temporally; compounding the issue, the growth to pebble sizes and larger renders a large fraction of the solids inemissive at sub-mm wavelengths and essentially undetectable \citep{Andrews20}. 
CO emission provides an alternative method to estimate the gas mass as CO has long been used as a tracer of \Htwo{} mass in molecular clouds.  However, the conversion of line emission to column density requires a prior knowledge of the temperature, CO distribution, and opacity to account for the fact that observations generally only sample the upper state column of one rotational state. The CO abundance relative to \Htwo{} remains relatively consistent throughout the dense ISM at approximately \(10^{-4}\) relative to \Htwo{} \citep{Frerking1982, Lacy94, Bergin17,Lacy17}. In protoplanetary disks, however, CO appears to be physically depleted and such depletion varies not only disk by disk \citep{Ansdell16, Miotello17, Long17}, but radially \citep{Zhang19,Zhang20_hd163}, and possibly temporally due to chemical and physical evolution \citep{Reboussin15, Boseman18, Cleeves18,  Eistrup, Krijt18, Schwarz19, Zhang20_evolution}. All of these factors complicate the use of CO to provide a reliable estimate of the total gas mass.

HD is a promising mass tracer, as the ratio of HD/\Htwo{} is well calibrated via measurements of the atomic D/H ratio \citep{Linsky98}.   However the conversion of emission of HD to mass is strongly temperature sensitive for typical temperatures between 20--50 K; this is due to the first rotational state (J = 1-0) having an \(\rm E_{\rm up}\) equal to 128.5~K above the ground \citep{Bergin13, Bergin17, Kama20_real}. A well constrained disk mass derived using HD, can only be determined if an accurate thermal structure is available  \citep{Trapman17, Kama20_real}. 



Given the wide reaching impact of the thermal structure of a protoplanetary disk, it has long been a property sought after. Before the advent of the Atacama Large (sub-)Millimeter Array (ALMA), the spectral energy distribution (SED) and spatially unresolved gas observations provided the best available constraints for most disks. One of the first attempts to uncover a thermal structure with single dish observations was \citet{vanZadelhoff01}, which utilized isotopologues of CO, HCO$^{+}$, and HCN. By measuring the ratios of flux, they extrapolated densities and temperatures from which that flux originated. But with unresolved observations, spatial information becomes degenerate as each of the isotopologues can be sensitive to different radial and vertical layers in the disk \citep{Gorti11,Zhang17,Woitke19}. Spectrally resolved observations of isotopologues have already been illuminating, shedding light onto the thermal profiles of a handful of disks, i.e. \citet{Fedele16}, which utilized both spectrally resolved and unresolved lines of high-$J$ CO to extract thermal profiles of disks. While spatially unresolved, the spectrally resolved observations still can extract spatial information from the Keplerian rotation, if the disk inclination is known.

Spatially resolved observations allow for measurements of column densities and temperature information at different disk radii \citep{Bruderer12,Rosenfeld13,Schwarz16,Zhang17,Pinte18} providing finer constraints in a forward modeling approach. \citet{Kama16} was one of the first to derive a thermal structure using a spatially resolved line, using CO 3-2 towards TW Hya. This resolved line in addition to much of the CO ladder produced improvements to the understanding of the TW Hya thermal profile. Towards IM Lup, \citet{Pinte18} observed the 2-1 transition of \CO{}, \thCO{}, and \CetO{} and directly extrapolated T$_{\rm gas}$ from spatially resolved optically thick lines, independent of any modeling. 

The thermal structure of TW Hya has been probed using its SED in concert with a large array of spatially unresolved lines including various transitions of CO, HCO$^+$, and H$_2$O \citep{vanZadelhoff01,Gorti11,Zhang17,Woitke19}. Often these thermal profiles were empirically derived using these observations. With new thermo-chemical models that take into account radiative transfer and chemical evolution, the disk thermal structure can be derived using a forward modeling approach by reproducing observations \citep{Woitke19} including spatially resolved lines providing finer insight into radial structure \citep{Kama16}. The goals of this paper are to lay out a new approach to uncover a gaseous protoplanetary disk thermal profile by utilizing multiple spatially-resolved observations. Using a thermo-chemical code, we aim to reproduce seven spatially-resolved CO line observations towards the class II, nearby  \citep[60pc;][]{Gaia18, Bailer-Jones18}, face-on TW Hya disk. It is worth noting that we do not seek to reproduce scattered light nor continuum observations. While these observations provide additional constraints on the temperature structure (more directly on the dust temperature and indirectly on gas temperature), a detailed dust model would require further assumptions on how dust properties vary with radius \citep{Huang18}, as well as assumptions on dust scattering, which add uncertainty in dust temperature determinations \citep{Zhu19}. Such simulated observations would be non-trivial with our current model, and the goal of this paper is to uncover the broader gaseous thermal profile and mass. While continuum and scattered light images provide insight into deviations from a smooth density distribution, that information may not translate into an effect on the gas density and temperature.

The present study uses data from the TW Hya Rosetta Stone Project (PI: Cleeves) along with other archival data sets. 
The complete set of observations, including the CO and isotopologue maps from \citet{Schwarz16} and \citet{Huang18}, span a wide range of optical depths, thereby tracing both the vertical and radial temperature simultaneously when observations are spatially resolved, in the case for optically thick molecules. Molecules that turn out to be optically thin will provide column density information. Additionally, we seek to reproduce the HD flux measured with \textit{Herschel} and the SED of the disk.

We detail our ALMA observations in Section \ref{sec:obs} and our modeling procedure in Section \ref{sec:RAC2D}. In Section \ref{sec:results} we explore our best fit thermal model to investigate possible model degeneracies. Comparisons to other models and the future of HD observations are explored in Section \ref{sec:A+D}, and we summarize our findings in Section \ref{sec:Conclusions}.

\section{Observations} \label{sec:obs}
We present new observations of \CO{} 2-1 and \CetO{} 2-1 taken as a part of the Rosetta collaboration. Archival ALMA data were also used for \CO{} 3-2, \CetO{} 3-2, 6-5, and \thCO{} 3-2, 6-5. The archival HD 1-0 flux measurement from \textit{Herschel} and SED fluxes from the literature were also used.    A summary of all observations can be found in Table \ref{ALMA}. Using \textsc{bettermoments}\footnote{https://github.com/richteague/bettermoments} \citep{Teague18}, moment zero maps were calculated for each of the CO observations described below and are presented in Figure \ref{obs}. Each moment zero map was then azimuthally averaged producing the radial emission profiles.  These radial profiles provided the constraints that we used below to determine the 2D thermal structure of TW Hya. 

\subsection{CO Observations from TW Hya Rosetta Project}
We report new observations of  \CO{} 2-1 and \CetO{} 2-1 as part of the program 2016.1.00311.S (PI Cleeves). The compact observations (baselines down to 15 m) were obtained on December 16, 2016 in configuration C43-3 for a total on-source integration time of 81 minutes. The extended observations (baselines up to 1124 m) were carried out on May 5 and 7, 2017 in C43-6 with an on source integration time of 25 minutes. The data were calibrated by the CASA pipeline \cite{McMullin07}. For the extended observations, J1058+0133 and J1037-2934 were used as bandpass calibrators, J1107-4449 as the flux calibration, and J1037-2934 as the phase calibrator. For the compact observations, J1037-2934 was used for bandpass, flux, and phase calibration. We performed one additional round of phase self calibration on each of the extended and compact observations independently using CASA version 4.5.0. For the self calibration we adopted a solution interval of 30 seconds and averaged both polarizations. Spectral windows were not averaged in the self calibration step, and the solutions were mapped to the individual spectral windows. 

Each data cube was first centered to the same central coordinate, then the continuum was subtracted from the line observations in the uv plane using the CASA routine \textit{uvcontsub}. The continuum subtracted long and short baselines were then combined using the method \textit{concat} where the spectral windows that excluded the lines were manually specified. The images were produced in CASA version 4.6.12 using \textit{tclean}, with Briggs weighting and a robust parameter of 0.5. The final spectral resolution of CO and \CetO{} were 246.08 kHz and 73.24 kHz respectively. The \CetO{} restoring beam had FWHM dimensions of \(0.''40 \times 0.''30\) and \CO{} had \(0.''37 \times 0.''28\).

\subsection{Archival CO Data}
The \CO{} $J$ = 3-2 analysis and imaging procedures are reported in \citet{Huang18} and the \thCO{} $J$ = 3-2, 6-5 and \CetO{} $J$ = 3-2, 6-5 observations are described in \citet{Schwarz16}, and not repeated here. We utilize their final data reduction and images.

\subsection{Spatially Unresolved Observations: \\ HD Flux and SED}

HD J=1-0 was observed using \textit{Herschel} and the observed total integrated flux  was \(6.3 (\pm0.7)\times 10^{-18}\) \(\textrm{W/m}^{2}\) \citep{Bergin13}. The line is spatially unresolved, thus we only use the integrated line flux to compare the models.

TW Hya is one of the most thoroughly observed protoplanetary disks in the literature, with a well-characterized spectral energy distribution (SED). The main purpose of the SED is to constrain the thermally coupled small dust population. Data points for the SED were taken from \citet{Cleeves15}, which in turn used individual photometric measurements from the literature \citep{Weintraub89, Weintraub00, Mekkaden98, Cutri03, Hartmann05, Low05, Thi10, Andrews12, Qi04, Wilner00, Wilner03}.

\section{Modeling: RAC2D} \label{sec:RAC2D}

We create a thermo-chemical model of the TW Hya disk that takes into account the gas and dust structure while simultaneously computing the temperature and chemical structure throughout the disk over time. Simulated observations are derived from the output model and compared to spatially-resolved line observations, as well as the SED and HD line flux. We use the time-dependent thermo-chemical code RAC2D \citep{Du14}. A brief description of the physical model follows; a higher-level detailed description of the code can be found in the aforementioned paper.

\subsection{Physical Structure}\label{structure}

Our model consists of three mass components: gas, a small dust (\(\leq\) \micron) population, and a large dust (\(\leq\) mm) population. Each population is described by a global surface density distribution \citep{Lynden-Bell74} which has been widely used in modeling protoplanetary disks:

\begin{equation}
\Sigma(r)=\Sigma_{c}\left(\frac{r}{r_{c}}\right)^{-\gamma}\exp{\left[-\left(\frac{r}{r_{c}}\right)^{2-\gamma}\right]}
\label{surfden}
\end{equation}

\noindent \(r_{c}\) is the characteristic radius at which the surface density is \(\Sigma_{c}\) and \(\gamma\) is the power-law index that describes the radial behavior of the surface density.  Each dust population follows an MRN grain size distribution \(n(a) \propto a^{-3.5}\) \citep{Mathis77} . The gas and small dust are spatially coupled and exist out to 200 AU \citep{vanBoekel17, Huang18} and possess a scale height that is related to a critical radius (defined below).  We assume that the large dust population is settled in the midplane and extends less than 100 AU \citep{Andrews10}. This reduced scale height approximates and represents the impact of dust evolution and pebble sedimentation to the midplane \citep{Birnstiel12, Krijt16}. The specific values for our TW Hya model are constrained by observation and detailed in the next section.

\begin{figure}
    \centering
    \includegraphics[scale=.4]{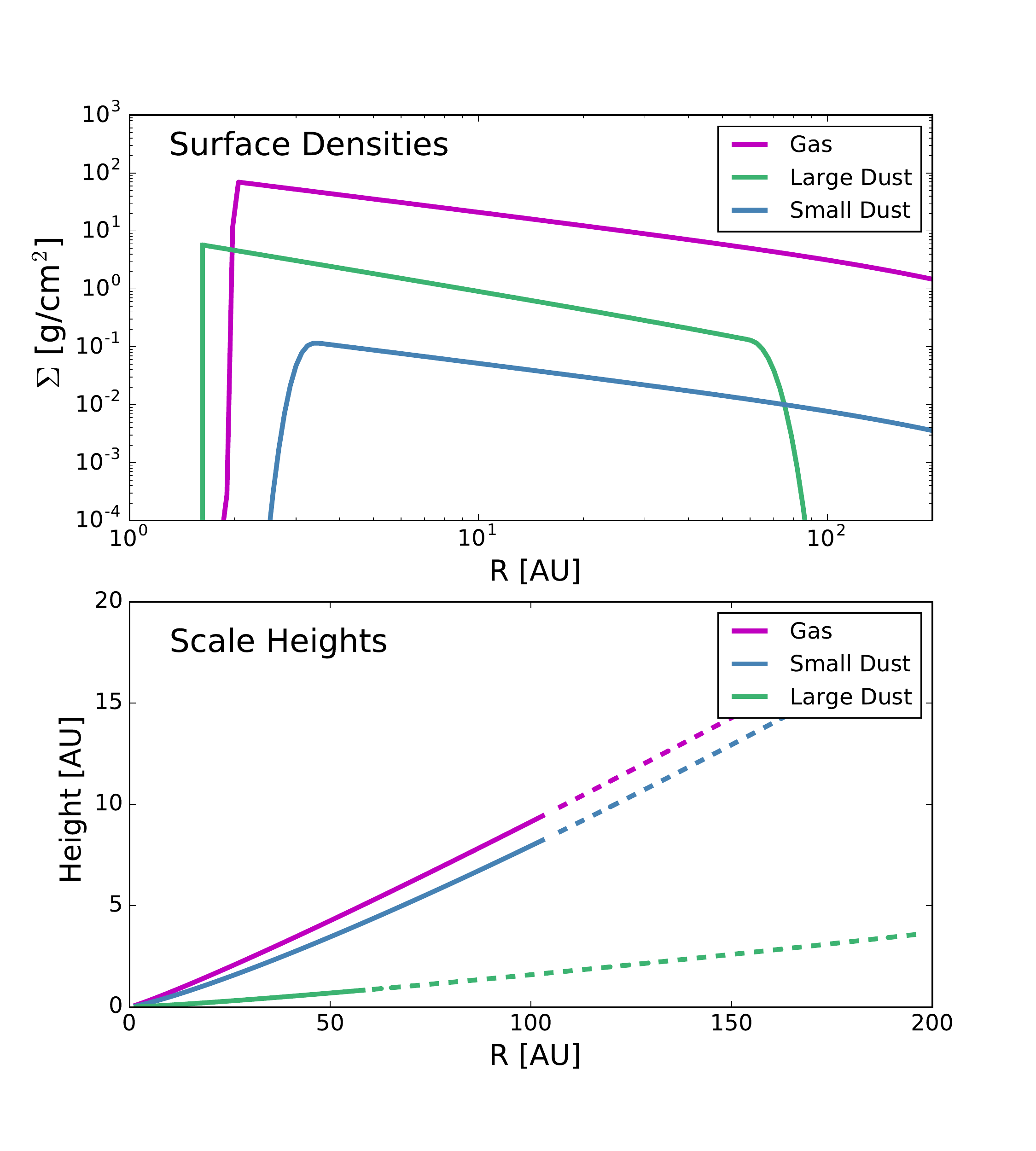}
    \caption{The surface density (top) and the scale heights (bottom) of each component of our disk model. The exponential taper seen added to the surface densities at 60 AU for the large dust and 104 AU for the small dust and gas components. The location at which dashed line starts for the scale height also corresponds to where the exponential taper for that population begins, corresponding to a significant decrease in density.}
    \label{fig:density_scaleheight}
\end{figure}

\begin{figure}[]
    \centering
    \includegraphics[scale=.6]{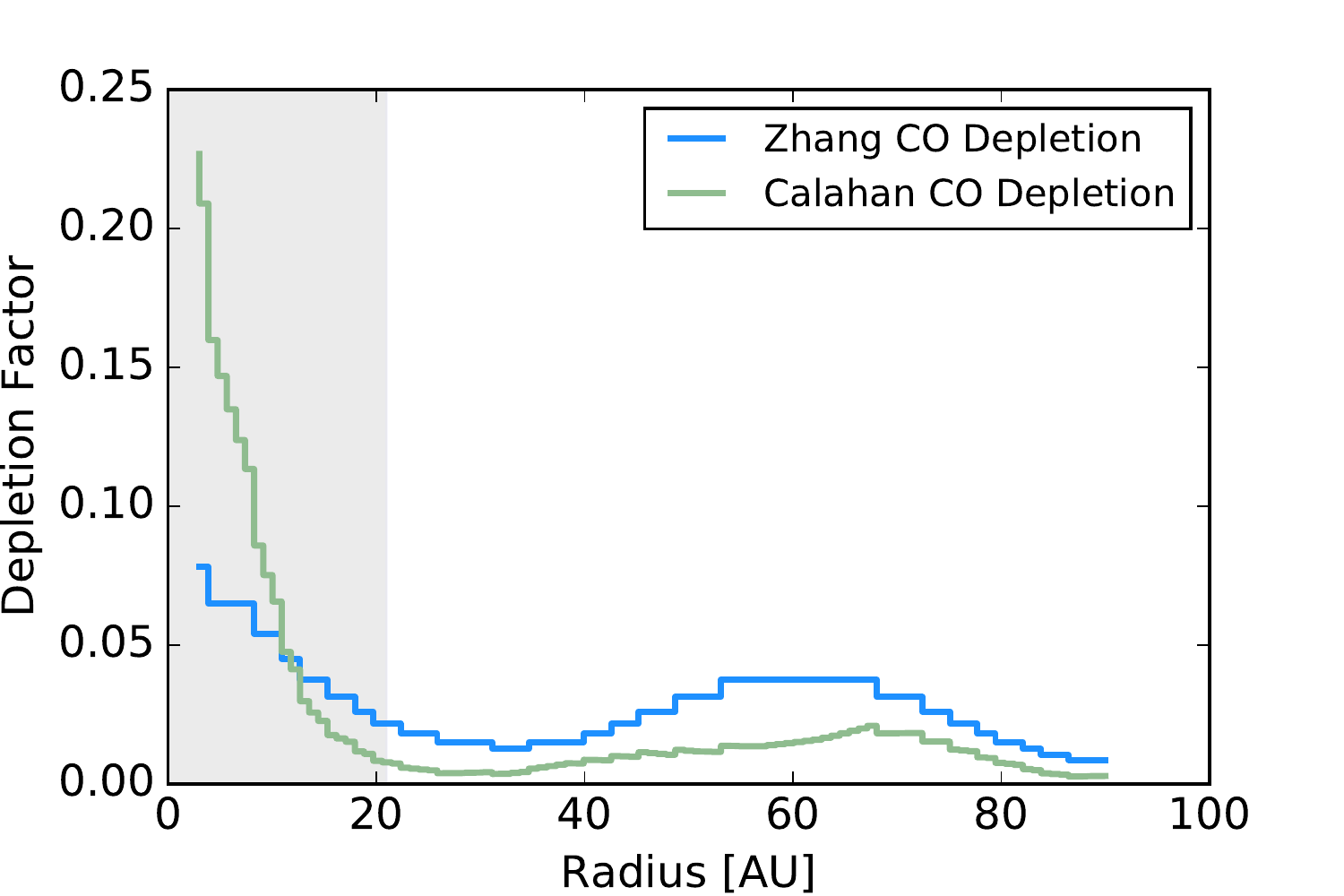}
    \caption{ The final radial depletion profile (green) as motivated by the depletion profiles derived in \cite{Zhang19} (blue). CO was depleted radially according to the above factor at the beginning of the thermo-chemical calculation. The transparent grey corresponds to the beamsize of the \CetO{} 2-1 observation. }
    \label{COdep}
\end{figure}

RAC2D is equipped with an exponential taper added to the surface density profile both in the inner and outer disk regions. This allows for more flexibility in the position of the exponential taper. If \(r_{c}\) is defined to be larger than the gas component (which would result in a flatter disk) an exponential taper can still be added. If we were to only use Equation 1, and required a large \(r_{c}\) value, there would be an abrupt cut-off at the user-defined outer radius. \(r_{exp}\) defines at what radius the exponential taper begins, and \(r_{s}\) describes the strength of that taper:
\begin{center}
\(\Sigma'(r) = \Sigma(r) e^{(r-r_{\rm exp_{\rm in}})/r_{s_{in}}}\), \textrm{if}   \( r < r_{\rm exp_{\rm in}}\)
\end{center}
\begin{center}
\(\Sigma'(r) = \Sigma(r) e^{-(r-r_{\rm exp_{\rm out}})/ r_{\rm s_{out}}}\), \textrm{if}   \( r > r_{\rm exp_{\rm out}}\)
\end{center}

\noindent A density profile for the gas and dust populations is derived from the surface density profile and a scale height:
\begin{equation}
    \rho(r,z) = \frac{\Sigma(r)}{\sqrt{2\pi}h(r)} \exp{\left[-\frac{1}{2}\left(\frac{z}{h(r)}\right)^{2}\right]}
\end{equation}

\begin{equation}    
    h=h_{c}\left( \frac{r}{r_{c}}\right)^{\Psi}
\end{equation}

\noindent Where \(h_{c}\) is the scale height at the characteristic radius, and \(\Psi\) is a power-law index that characterizes the flaring of the disk structure. The final surface densities and scale heights for each component in our model is seen in Figure \ref{fig:density_scaleheight}. In the surface density plot, we show that although all of our components (gas, small dust, and large dust) have a $\rm r_{c} $ value equal to 400 AU, they exponentially drop off at their given $\rm r_{exp_{out}}$.

\subsection{Dust and Gas Temperatures}
After initializing RAC2D with a model density structure for each population, the code computes a dust and gas temperature. The determination of dust and gas temperature is an iterative process, allowed to change over time due to the evolving chemical composition. For the gas-phase chemistry, we adopt the reaction rates from the UMIST database \citep{Woodall07}, with additional rates considering the self-shielding of CO, \Htwo{}, \(\textrm{H}_{2}\textrm{O}\), and OH, dust grain surface chemistry driven by temperature, UV, cosmic rays, and two-body chemical reactions on dust grain surfaces \citep[see references given by][]{Du14}. Chemical processes also provide heating or cooling to the surrounding gas. That, along with stellar and interstellar radiation drive the thermal gas structure. Our study explores models that account for 1 Myr of chemical evolution. Although the age of TW Hya is estimated to be 5-10 Myr \citep{Debes13}, our assumed depletion profile for CO (Section \ref{depletion}) encapsulates the effects of earlier chemical evolution of CO.  We are essentially modeling the thermal physics and chemistry after that evolution has occurred. Finally, at the end of a given run we extract the dust and gas thermal profiles and SED profile. 

Simulated line images for CO, \thCO{}, \CetO{}, and HD are necessary for our comparison to observations. We do not model isotopologue fractionation in this chemical network, as fractionation of CO is not significant in a massive disk like TW Hya \citep{Miotello14} . Thus, we compute \thCO{} and \CetO{} abundances based on ISM ratios of CO/\thCO{} = 69 and CO/\CetO{} = 570 \citep{Wilson99}. We then apply the depletion profile to each CO line. The HD abundance (see Table \ref{chemAbundance}) remains constant. Given these abundances and the local gas temperature, RAC2D computes line images using a ray-tracing technique. We then manually convolve these simulated observations with the corresponding ALMA beam to make direct comparisons to data. Our HD 1-0 observations are spatially and spectrally unresolved, thus to recreate unresolved integrated flux measurements, we convolve our model HD line over a gaussian corresponding to the velocity resolution of the \textit{Herschel} PACS instrument: $\sim$300  ${\rm km\,s^{-1}}$ \citep{Poglitsch10}.

The SED is created by RAC2D by counting the number of photons over a range of frequencies that interact with the disk and would fall within a given range of sight lines. Thus photons directly emitted from the star are not accounted for, a design motivated by computational efficiency. To create the final SED the input stellar blackbody of TW Hya (stellar mass = 0.8 \Msun{} and T$_{\textrm{eff}}$ $\approx$ 3400K) is added to the result produced by RAC2D.

\renewcommand{\arraystretch}{1}
\begin{deluxetable}{cc}
    \tablewidth{0pt}
    \tablecolumns{2}
    \tablecaption{Initial Chemical Abundances for Final Model}
    \tablehead{
    &\colhead{Abundance Relative to Total}\\
    &\colhead{Hydrogen Nuclei}}
     \startdata
      \Htwo & \( 5 \times 10^{-1}\) \\
      HD & \(2 \times 10^{-5}\) \\
      He & 0.09\\
      CO* & \(1.04 \times 10^{-4}\)\\
      N & \(7.5 \times 10^{-5}\)\\
      \(\textrm{H}_{2}\textrm{O (ice)}\) & \(1.8 \times 10^{-4}\)\\
      S & \(8 \times 10^{-8}\)\\
      Si+ & \(8 \times 10^{-9}\)\\
      Na+ & \(2 \times 10^{-8}\)\\
      Mg+ & \(7 \times 10^{-9}\)\\
      Fe+ & \(3 \times 10^{-9}\)\\
      P & \(3 \times 10^{-9}\)\\
      F & \(2 \times 10^{-8}\)\\
      Cl & \(4 \times 10^{-9}\)\\
     \enddata
     \tablecomments{*CO has an imposed radial depletion profile as shown in Figure \ref{COdep}}
     \label{chemAbundance}
\end{deluxetable}

\begin{deluxetable*}{cccl}
    \tablewidth{0pt}
    \tablecolumns{2}
    \tablecaption{Gas and Dust Population Parameters: Ranges and Initial Values}
    \tablehead{
    &\multirow{2}{*}{Gas} & Small Dust  &\multirow{2}{*}{Definition} \\
    &&(\(5 \times 10^{-3}\) - 1 \(\micron\))&}
     \startdata
     Mass (\Msun) & 0.01 - [0.05] - 0.06 & \( 2.5 \times 10^{-5} - [1 \times 10^{-4}] - 5 \times 10^{-4}\)  & Total Mass\\
     \(\Psi\) & 0.8 - [1.2] - 1.6 & 0.8 - [1.2] - 1.6 &  Flaring Parameter\\
     \(\gamma\) & 0.5 - [.75] - 1.1 & 0.5 - [.75] - 1.1 &  Surface Density Power-Index\\
      \(\textrm{h}_{c}\) & 8.4 - [42] - 84 & 8.4 - [42] - 84 &  Characteristic Height \\
     \(r_{{\rm exp}_{\rm out}}\) & 60 - [104] - 200  & 60 - [104] - 200 &  Exponential Drop-off Radius in the Outer Disk \\
    \(r_{{\rm s}_{\rm out}}\) & 70.7  & 70.7 &  Exponential Drop-off Strength in the Outer Disk \\
     \(r_{{\rm exp}_{\rm in}}\) & 2  & 3.5 &  Exponential Drop-off Radius at the Inner Disk \\
     \(r_{{\rm s}_{\rm in}}\) & 0.1  & 0.5 &  Exponential Drop-off Strength at the Inner Disk \\
     \(r_{in}\) (AU) & 0.1 & 0.5 &  Inner Radius Cut-off \\
     \(r_{out}\) (AU) & 200 & 200 &  Outer Radius Cut-off \\
     \(\textrm{r}_{c}\) & 400 & 400 &  Characteristic Radius\\
     \enddata
     \tablecomments{This table shows the range of parameter space explored in each variable of our thermo-chemical model, with the initial value in brackets. Since the parameter values for the larg dust do not significantly impact the thermal structure, we do not explore variations in the equivalent parameters in the large dust component and keep these fixed to the Zhang et al. values (see Table 4).  All length values are in units of AU. \label{paramsrange}}
     \label{paramsrange}
\end{deluxetable*}

\begin{deluxetable*}{cccc}
    \tablewidth{0pt}
    \tablecolumns{2}
    \tablecaption{Gas and Dust Population Parameters: Final Model Values}
    \tablehead{
    &\multirow{2}{*}{Gas} & Small Dust & Large Dust \\
    &&(\(5 \times 10^{-3}\) - 1 \(\micron\))&  (\(5\times 10^{-3}\) - \(1 \times 10^{3}\) \(\micron\))}
     \startdata
     Mass (\Msun) & 0.025 & \( 1.0 \times 10^{-4}\) & \( 4.0 \times 10^{-4} \) \\
     \(\Psi\) & 1.1 & 1.2  & 1.2 \\
     \(\gamma\) & 0.75  & 0.75  & 1.0 \\
     \({\rm h}_{c}\) & 42 & 42 & 8.4 \\
     \(r_{{\rm exp}_{\rm out}}\) & 104  & 104 & 60 \\
     \(r_{{\rm s}_{\rm out}}\) & 70.7  & 70.7 & 10 \\
    \(r_{{\rm exp}_{\rm in}}\) & 2  & 3.5 & N/A \\
     \(r_{{\rm s}_{\rm in}}\) & 0.1  & 0.5 & N/A \\
     \(r_{in}\) & 0.1 & 0.5 & 1 \\
     \(r_{out}\) & 200 & 200 & 200  \\
     \({\rm r}_{c}\) & 400 & 400 & 400 \\
     \enddata
     \tablecomments{Final values of the TW Hya model that reproduces the CO, HD, and SED observations. All length values are in units of AU.}
     \label{params}
\end{deluxetable*}

\begin{figure*}
    \centering
    \includegraphics[scale=.475]{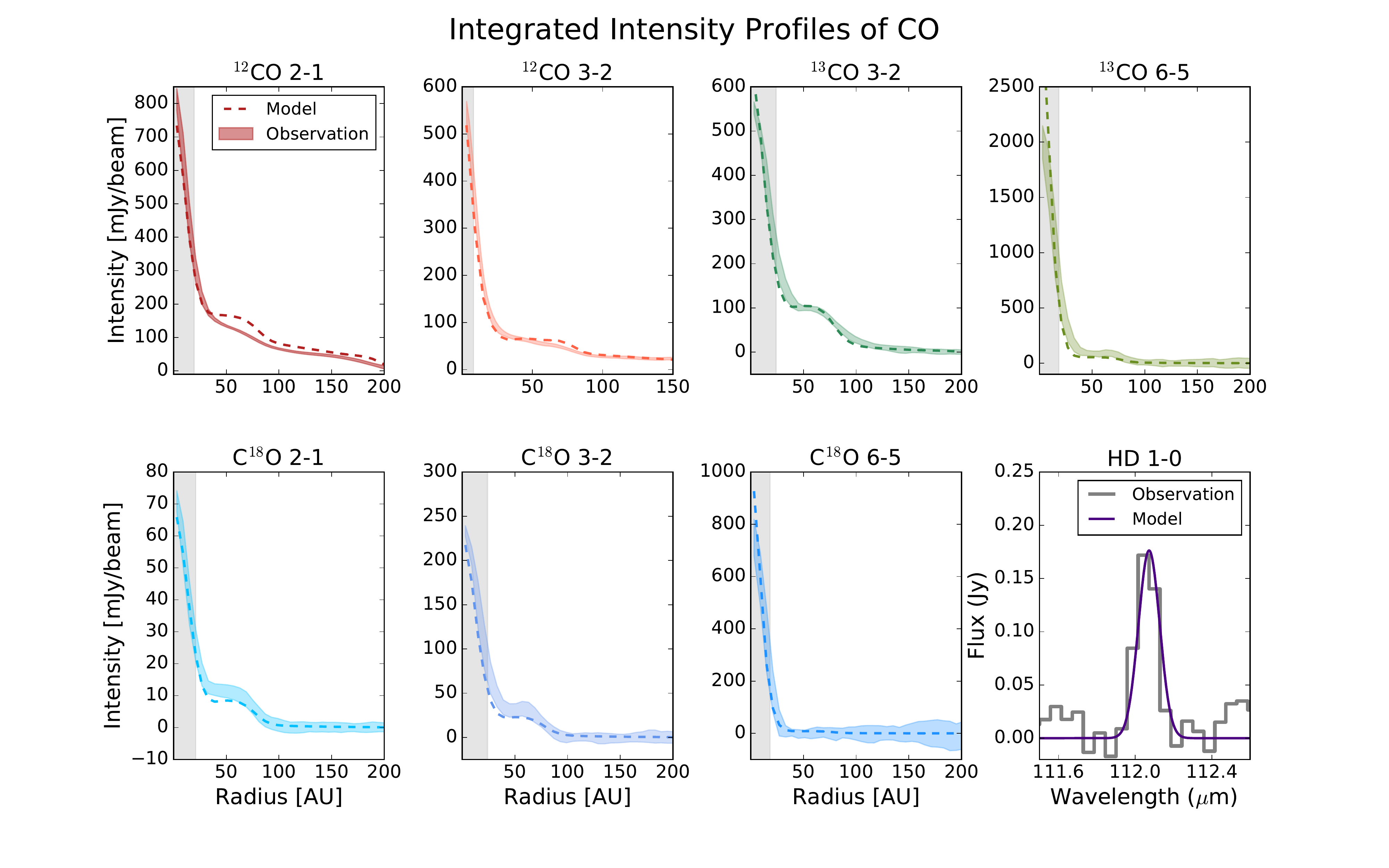}
    \caption{Integrated radial intensity profiles of CO and its isotopologues \thCO{} and \CetO{}. The solid lines are from observations from ALMA (see Table \ref{ALMA}) and the dashed lines are from our best-fit model, and each observation's beam is shown in the filled region. The thermal profile that produced this best match to the observation is shown in Figure \ref{temperature}. }
    \label{COprof}
\end{figure*}

\subsection{Initial Parameter Setup of TW Hya}\label{depletion}

We set out to create a model which, given certain initial parameters and constraints for the disk physical state, produces a gas thermal profile that can reproduce the spatially resolved CO observations and other constraints. Our initial parameters are taken from \cite{Zhang19}  who compiled the most up-to-date constraints on disk parameters from the literature: \cite{ Brickhouse10, Andrews12, vanBoekel17, Gaia18,Huang18}.  A table of source information is found in Table 3 of \cite{Zhang19}. For initial chemical abundances, we adopt values given in Table \ref{chemAbundance} motivated by \citet{Du14} and \citet{Zhang19}. The disk inclination used is 6$^{\circ}$, as past studies of TW Hya tend to use a range of inclination values between 5-7$^{\circ}$ \citep{Qi04,Andrews12,Huang18} 

In this study, we begin with a radial CO abundance profile motivated by \citet{Zhang19}, which calculated two depletion profiles, one for \thCO{} and one for \CetO{}. They were calculated to fit the isotopologue observations. The depletion profile dictates how much CO should be depleted from the expected total value, including both gas and solid state  ($\sim 10^{-4}$ for $^{12}$CO) at each radius. The difference in the observed depletion profiles between the isotopologues can be understood by differential effects between \thCO{} and \CetO{} such as fractionation and isotopic selective self-shielding \citep[e.g.][]{Visser09,Woods09,Miotello14, Miotello16}. Note that the impact of isotope selective self-shielding is much smaller than the overall removal of CO from the surface layers \citep{Du14}. Hence, our initial depletion profile was the average of the \thCO{} and \CetO{} observed depletion, and is shown in Figure \ref{COdep}. In this study, we implemented the CO depletion at the start of the chemical and thermal evolution, as the lack of CO will affect the temperature structure. We found that with the \cite{Zhang19} CO depletion profile, our simulated radial profiles did not reproduce the \CetO{} 2-1 and \thCO{} 2-1 profiles, in particular. This is due to both the chemical processing over the timescale of 1 Myr and the change in temperature at which each of these transitions emit. Reducing the chemical time to 0.01 Myr removed the effect of chemical processing, which should be encompassed through the implemented CO depletion. Further iterations on the radial CO depletion profile were made until we reached a model that reproduced the \CetO{} 2-1 observations within the margins of uncertainty. Our final CO depletion profile is shown in comparison to that of \cite{Zhang19} in Figure \ref{COdep}. The largest difference between the two exists within the inner 20 AU where we found that we did not require as extreme a CO depletion factor. Beyond 20AU the final CO depletion profile is on average 2.5 times less than what was derived in \cite{Zhang19}

\begin{figure}
    \centering
    \includegraphics[scale=.35]{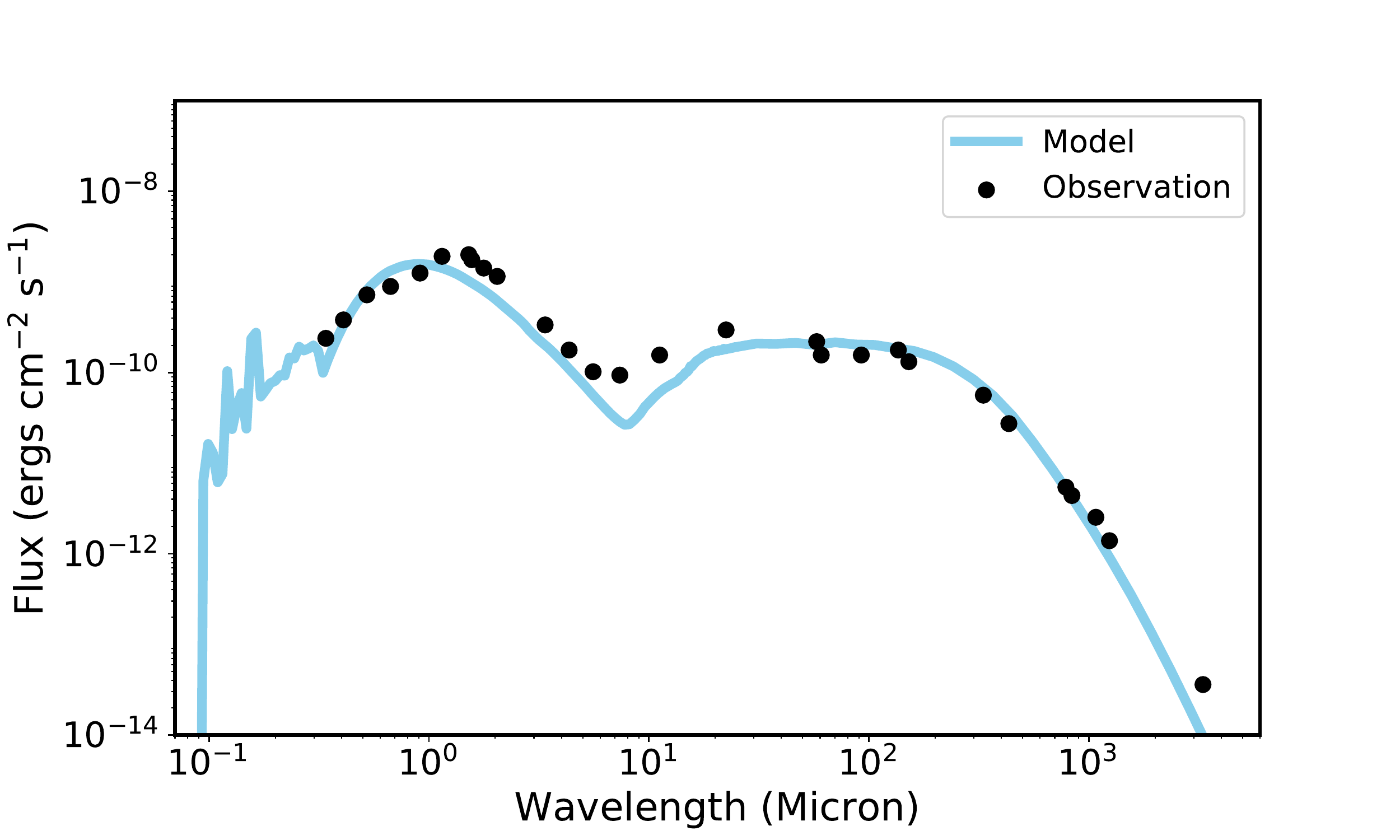}
    \caption{A comparison of the photometry data that comprises TW Hya's SED and the modeled SED from RADMC3D based off of the gaseous and dust density and temperature structures from RAC2D.} 
    \label{SEDfigure}
\end{figure}
\subsection{Parameter Exploration}

This study relies on a suite of CO line observations, as each line provides information about a slightly different vertical layer within the disk \citep{Gorti11,Zhang17,Woitke19}. As the CO observations were spatially resolved, the quality of model was assessed by a comparison of the radial integrated intensity profiles. We additionally seek to match the HD flux and the SED.

As a first step we set up a grid of models to explore the full parameter space. The parameter space and initial model parameters from \cite{Zhang19} are shown in Table \ref{paramsrange}.  Initially, we explored gas mass, small dust mass, \(\Psi\), \(\gamma\), and radii and height values within ranges supported by previous observation and modeling work. Each parameter is explored using values above and below the initial value. For the gas mass, our upper limit was 0.06 \(\textrm{M}_{\odot}\) which is the upper limit quoted in \cite{Bergin13} and the lower limit was 0.01 \Msun{}, which is similar to some earlier estimates of TW Hya's gas mass \citep{Weintraub89a, Thi10}. The small dust mass range was motivated by the SED, limiting the range from factors of 2$\times$ greater than our initial value to \(\sim\)3$\times$ less. Both extremes produce SEDs that do not match the observations, but values in between provide a reasonable match to the SED. The \(\Psi\) values explored vary from 0.8 to 1.3; both extremes affect the integrated CO emission profiles quite strongly. \(\gamma\) has an upper limit of 2 (see Equation \ref{surfden}), however, we explore smaller variations on gamma around an initial value of 0.85, thus using limits of 0.5 - 1.1. We did not explore values nearing the upper limit of two as past observational constraints on \(\gamma\) suggest the value for TW Hya is not much larger than 1.0 \citep{Kama16,vanBoekel17,Zhang17}. While this is a narrow range, it did provide enough information to understand the trends that come from changes in \(\gamma\). For \(r_{exp_{\rm out}}\), \(h_{c}\), and \(r_{c}\), the smallest value explored was that of the large dust value, and the largest value explored was double the initial model value. We did not explore the inner nor outer edge values of \(r_{s}\) (see Section \ref{structure}) nor the inner exponential radial cutoff \(r_{\rm exp_{\rm in}}\) as these values cannot be constrained with our CO observations. 
For each model we compare the model output CO isotopologue emission profiles as shown in Figure \ref{COprof} in search of the best fit.

\section{Results} \label{sec:results}

\begin{figure}
    \centering
    \includegraphics[scale=.42]{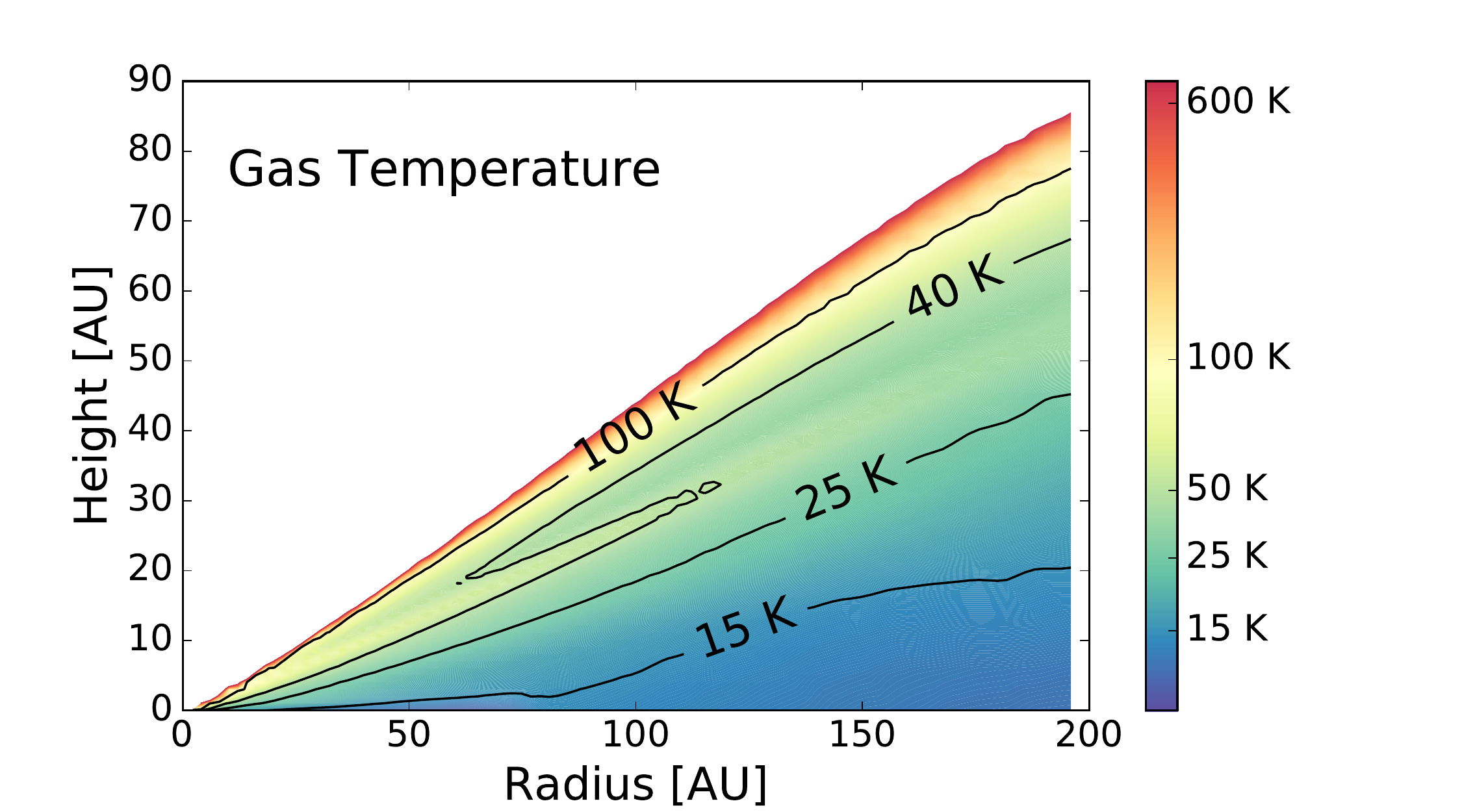}
    \caption{The final 2D thermal structure of the gas that reproduces seven resolved ALMA CO lines, HD flux, and the SED. The final temperature profile is supported up to scale heights from which \CO{} 2-1 emits from, since this transition has the highest optical depth}
    \label{temperature}
\end{figure}

\subsection{Description of Best-Fit Model}

We show the results of our best-fit model in Figure \ref{COprof}, \ref{SEDfigure}, and \ref{temperature}. The model agrees well with all seven resolved CO lines available along with the HD J=1-0 flux, and the SED profile (Figure \ref{SEDfigure}). The parameter values of the best-fit model are given in Table \ref{params}.

Our beginning thermal structure was from the TW Hydra disk model of \citet{Zhang19} (see Section \ref{depletion}). This initial model under-represents the CO flux in the inner \(\sim\)25 AU while over-representing the CO flux between 25 and 75 AU particularly in the \thCO{} J=3-2 and \CetO{} J=2-1 lines (see Figure \ref{fig:original}). The initial model HD line flux is \(4.6 \times 10^{-18} \) \(\textrm{W m}^{-2}\), which is 73\% of the observed value, falling outside of the quoted error of that flux measurement from \cite{Bergin13} \(6.3 (\pm 0.7) \times 10^{-18} \) \(\textrm{W m}^{-2}\). To solve these initial discrepancies, we explore the parameter space extensively (see more details  in the Appendix). We find that in order to match both CO and HD observations the \(\Psi\) values of the gas and small dust need to be slightly different (1.1 and 1.2, respectively). This difference creates a thin layer at the surface of the disk which is depleted of dust. The difference between the scale height of the gas and dust is ~1.2\% of the radial location (represented in Figure \ref{fig:density_scaleheight}). 

This thin layer depleted of dust is a crude way to represent the impact of coagulation and settling on the small dust population end of the assumed MRN  \citep{Dullemond04}. The changing values of \(\Psi\) and thus scale heights can be seen as a way of representing a modified MRN distribution. 
Although the grain size distribution is not directly explored in this study, we find that the distribution of small grains strongly influences the final thermal structure. This relative change in scale height between dust and gas appears to be the best method to significantly increase the intensity of the CO emission in the inner 25 AU. Gas in this thin upper layer is more effectively heated as UV radiation penetrates more easily when the dust component is reduced. This heats CO gas in slightly denser regions
and increases the CO emissivity.   

Summing all these effects together, our best-fit model has a total gas mass that is only half of that in \citet{Zhang19} model. This model predicts an HD flux of \(5.9 \times 10^{-18} \) \(\textrm{W m}^{-2}\), which is well within the uncertainty of the observed flux.  

\subsection{Best-fit SED}

The simulated SED as derived by our final TW Hya model was calculated using RADMC3D \citep{Dullemond12}. After the  temperature and physical distribution of the gas and dust were determined by RAC2D, we feed in those results including the stellar spectrum and dust opacities into RADMC3D.

Our model assumes a mass of small dust grains equal to \(10^{-4}\) \(\textrm{M}_{\odot}\) and a mass of the large dust grains to be \(4\times10^{-4}\) \(\textrm{M}_{\odot}\). Using these dust components, our model SED agrees reasonably well with the observed SED, see Figure \ref{SEDfigure}. TW Hydra also exhibits a strong silicate feature near 10~\micron{} which our dust model under predicts. Previous studies that have successfully modeled this feature by utilizing a special type of silicate within 4 AU \citep{Calvet02,Zhang13}. This is a scale much smaller than we aim to constrain, and is a small effect in comparison to the total dust opacity in this region. The SED beyond \(\sim\)8~\micron{} tests if we are correctly representing the dust population, as the long-wavelength portion of the SED is mostly affected by dust mass, size distribution, spatial distribution, and composition (a mixture of silicates and graphite). Overall, our dust model provides a good match to the observed SED. Since the small dust population has an effect on the thermal balance, the SED provides one of the key factors to constrain the HD emission and overall disk mass. 

\begin{figure*}
    \centering
    \includegraphics[scale=.47]{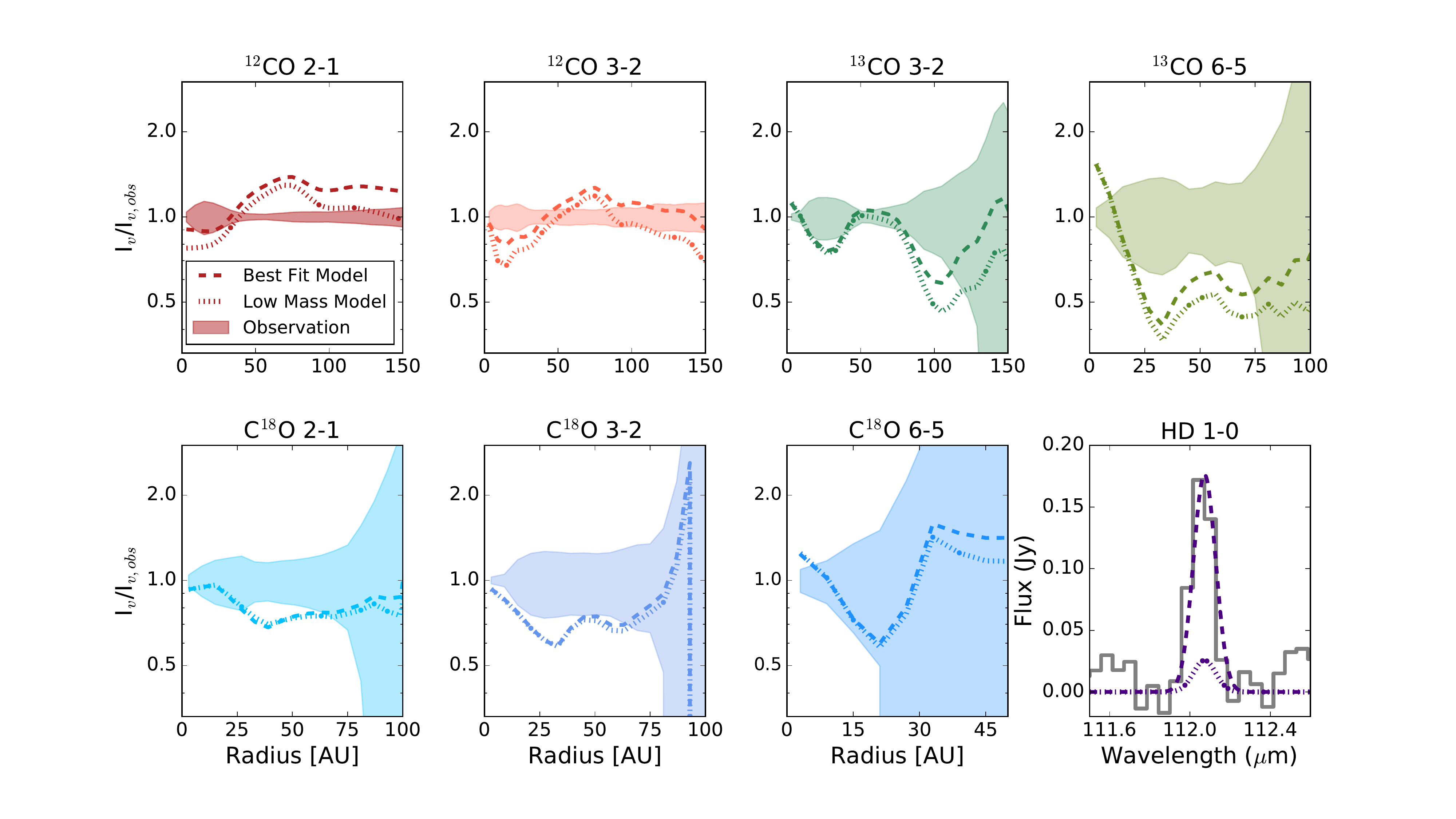}
    \caption{The radial intensity profiles of the observations, best fit model, and low mass model all normalized to the observations, along with the HD flux observations compared to the simulated HD flux in the low mass model. The observations are thus located at 1, and are the solid line while the models are dashed lines with the thick dashed line representing the best fit model (see Figure \ref{COprof}) and the low mass model (with high CO abundance) represented with the thinner dashed line. Errors of the integrated intensity line profile is shown in the shaded region, and is also normalized by the observed intensity. From this comparison, we can see that both the model with a gas mass of 0.023 \(\textrm{M}_{\odot}\) and 0.0023 \(\textrm{M}_{\odot}\) have nearly identical profiles. This also shows how close our model represents the observations. The majority of the time, the modeled flux is within a factor of two of the observed, with the only exceptions being when the observations drop quickly to receive zero flux and the model does not reach zero as quickly, however the model still is within the error. This happens especially with \CetO{} 6-5. While in the low mass model, the CO remains relatively similar, the HD flux drops significantly.}
    \label{fig:CompareCOdep}
\end{figure*}

\subsection{$^{13}C^{18}O$ Flux}

Using our final model, we seek to compare the observed $^{13}$C$^{18}$O flux towards TW Hya as presented in \citet{Zhang17}. We use line information for $^{13}$C$^{18}$O derived by HITRAN \citep{Rothman05}, formatted to be identical to line data from the Leiden Atomic and Molecular Database. We also use an abundance ratio \CO{}/\thCO{} = 40, as measured by \citet{Zhang17} for this specific observation.\footnote{We note that this is different than the assumed isotope ratio for the surface layers.  However, it is a direct measurement and is suggested by \citet{Zhang17} to be the result of carbon fractionation in the dense colder layers of the photodissociation region \citep[see discussion and results in][]{Rollig13}.}  The resultant flux is shown in Appendix Figure \ref{13C18O}. Overall, our model provides a good fit to the observations, underpredicting the $^{13}$C$^{18}$O flux by at most by 12 mJy km s$^{-1}$ beam$^{-1}$, or about a factor of two across the modeled flux emitting area. As the $^{13}$C$^{18}$O emission is highly centrally concentrated the location of largest discrepancy is in the central regions and also towards the observed asymmetry. This residual value is similar to the background noise flux. The asymmetry can be accounted for by simply the low signal/noise ratio for this observation.

\subsection{CO Mass/Abundance Degeneracy}

Our exploration of the parameter space shows that the CO line intensity is a degenerate result of the total gas mass and CO abundance. If the line is optically thin, the CO emission scales directly with its number density, $n_{\ce{CO}}$, where $n_{\ce{CO}}$ is $x_{\rm CO} \times$ $n_{\rm H_{2}}$. Even in the case of optically thick lines, as the gas density increases, CO emission becomes optically thick at a higher location of the disk atmosphere;  thereby tracing temperature of a warmer region. Thus there is a degeneracy between the abundance of CO and the density of \Htwo{} and hence the gas mass. One way to break this degeneracy is with an observed HD flux, as it is independent of CO abundance. 

To explore this, we create a low-mass model based on the parameters of our best-fit model, however in this model  we decrease the total gas mass by one order of magnitude while subsequently increasing in the CO abundance by the same factor. 
As shown in Fig.~\ref{fig:CompareCOdep} this low mass model produces indistinguishable CO radial profiles compared to these of our best-fit model.
However,  the line flux of HD (1-0) transition of the lower mass model goes down by more than a factor of five (see bottom right panel in Figure \ref{fig:CompareCOdep}). This degeneracy between CO abundance and gas mass has been reported in previous studies with less well-constrained temperature structures. \citep{Bergin13, Favre13, McClure16, Kama16, Trapman17}. Here we prove that the degeneracy cannot be broken even the thermal structure of the disk is well constrained. In short, CO lines alone are not sufficient to constrain the total disk mass and another independent mass tracer must be introduced, such as HD.   


\section{Analysis \& Discussion} \label{sec:A+D}

\begin{figure}
    \centering
    \includegraphics[scale=.4]{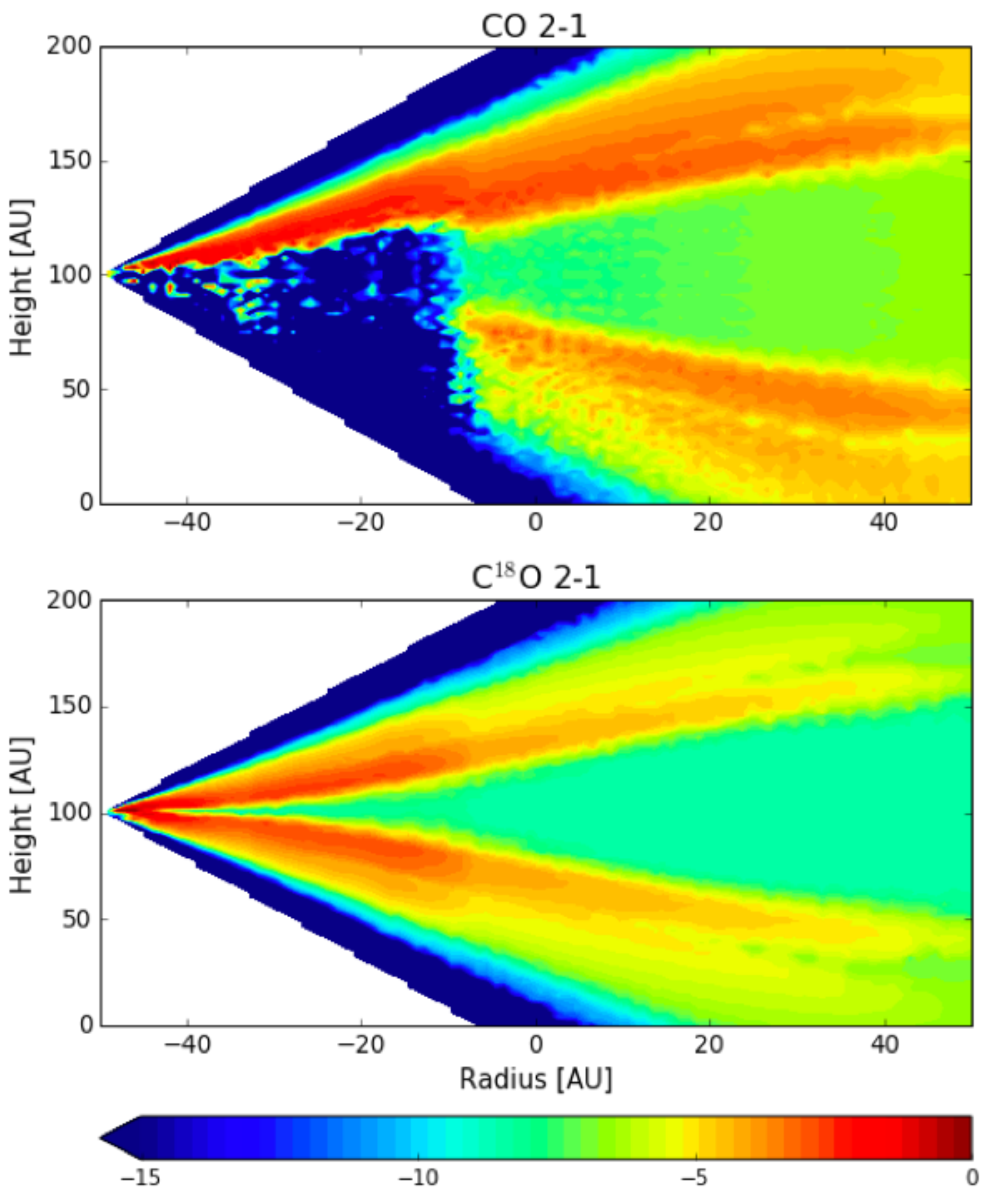}
    \caption{The emitting regions of \CO{} and \CetO{} 2-1 in our final model. Emitting regions for each molecule were calculated based on the final model, and were used to create Figure \ref{LineContri}, which encapsulates the regions in which most of each molecule is emitting from. }
    \label{lineContri_examples}
\end{figure}

\begin{figure*}
    \centering
    \includegraphics[scale=.6]{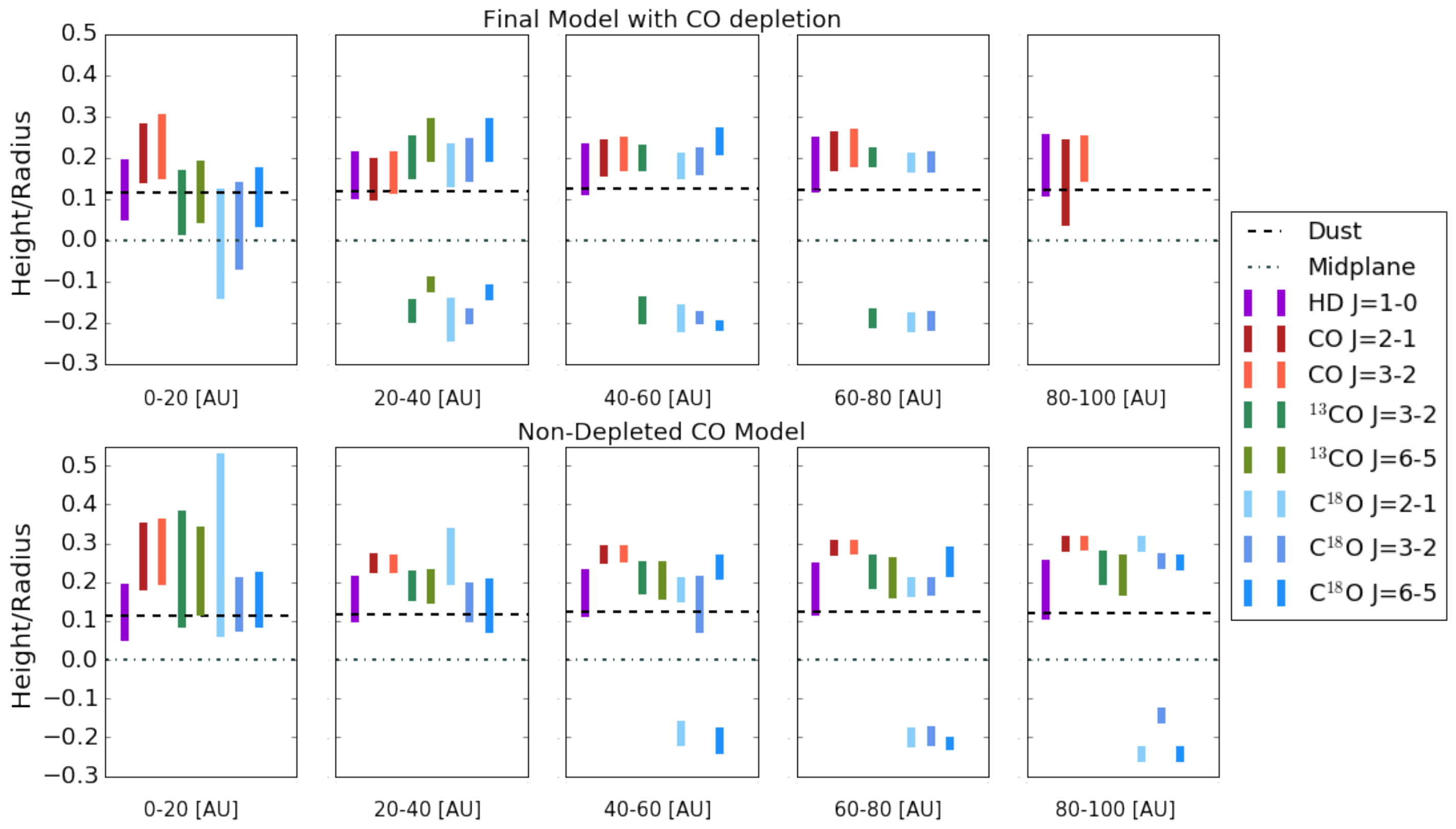}
    \caption{The average vertical emitting region of each line used in this study for our final model (top) and a version of that model without assuming CO depletion (bottom). For every 20 AU the average vertical emitting height was calculated. The top of each bar corresponds to the average surface which consists of 90\% of the flux. The bottom of the bar and below is 10\% of the total flux. Lines that have two emitting regions, one above the disk and below, are optically thin, thus observations of these molecules come from both the front and back of the disk. There is no emission in the midplane outside the CO snowline, thus for an optically thin emitter, their emitting regions were calculated twice. Once for above the midplane, then again but only from the back. The continuous \(\tau=1\) surface of the dust at the frequency of HD J=1-0 is shown via the dashed line. Examples of 2D contribution plots used to create this figure of \CO{} 2-1 and \CetO{} 2-1 are shown in Figure \ref{lineContri_examples}.}  
    \label{LineContri}
\end{figure*}


\subsection{CO Emitting Regions}\label{line_writing}

RAC2D calculates the emitting regions for each isotopologue; examples of \CO{} 2-1 and \CetO{} 2-1 are shown in Figure \ref{lineContri_examples}.  To allow for inter-molecular comparison, we then calculated the heights at which the majority of the emission originates.  This is presented in Figure \ref{LineContri} where we show a visualization of the average vertical cross-section of the disk from which each line is emitting based on our final model. For comparison, a non-CO depleted model is also shown. A dashed line representing the \(\tau = 1\) surface of the small dust at the frequency of the HD $J$=1-0 transition (\(\lambda \simeq 112\micron\)) is plotted as a reference. These two models represent two extremes of T Tauri disks, a high CO abundance (which corresponds to a CO/\Htwo{} ratio equal to that of the ISM) and a low CO abundance, using the relatively high CO depletion factors found in TW Hya. 

In the non-depleted case, \CO{} emits from a higher layer than all other isotopologues, while HD and \thCO{} and in some cases \CetO{}, probe similar depths. \CetO{} appears to be optically thin, especially past 60 AU; the \CetO{} surface brightness is therefore not a sufficient temperature tracer across a disk with similar properties as TW Hya. In this instance \CetO{} probes the total column density.  TW Hya is, however, is thought to exhibit severe CO depletions, thus CO and its isotopologues emission will have reduced optical depth (in comparison to the non-depleted model). In our final model, \CO{} 2-1, 3-2, and \thCO{} 3-2 emit from roughly an equivalent vertical depth as HD, showing that in this disk, observations of \CO{} 2-1, 3-2 and \thCO{} 3-2  best trace the layers where HD emission arises. Within radii ranging from 0 - 40 AU, \thCO{} 3-2 emission is optically thick, overlapping with the lower bound of the HD emitting layer. The \CetO{} and \thCO{} 6-5 become optically thin beyond radii of 20-40AU, thus their brightness temperature no longer constrains the disk surface temperature. Based on this result, the disk mass for TW Hya could have been accomplished using only the resolved observations of \CO{} 2-1, 3-2, and supplemented with \thCO{} 3-2 alongside HD. For future observations, multiple observations of CO are still required, at least down to the 3-2 levels, as CO depletion and gas mass are not known a priori. 

\subsection{Scattered Light Effects}\label{sec:effects}

\begin{figure*}
    \centering
    \includegraphics[scale=.5]{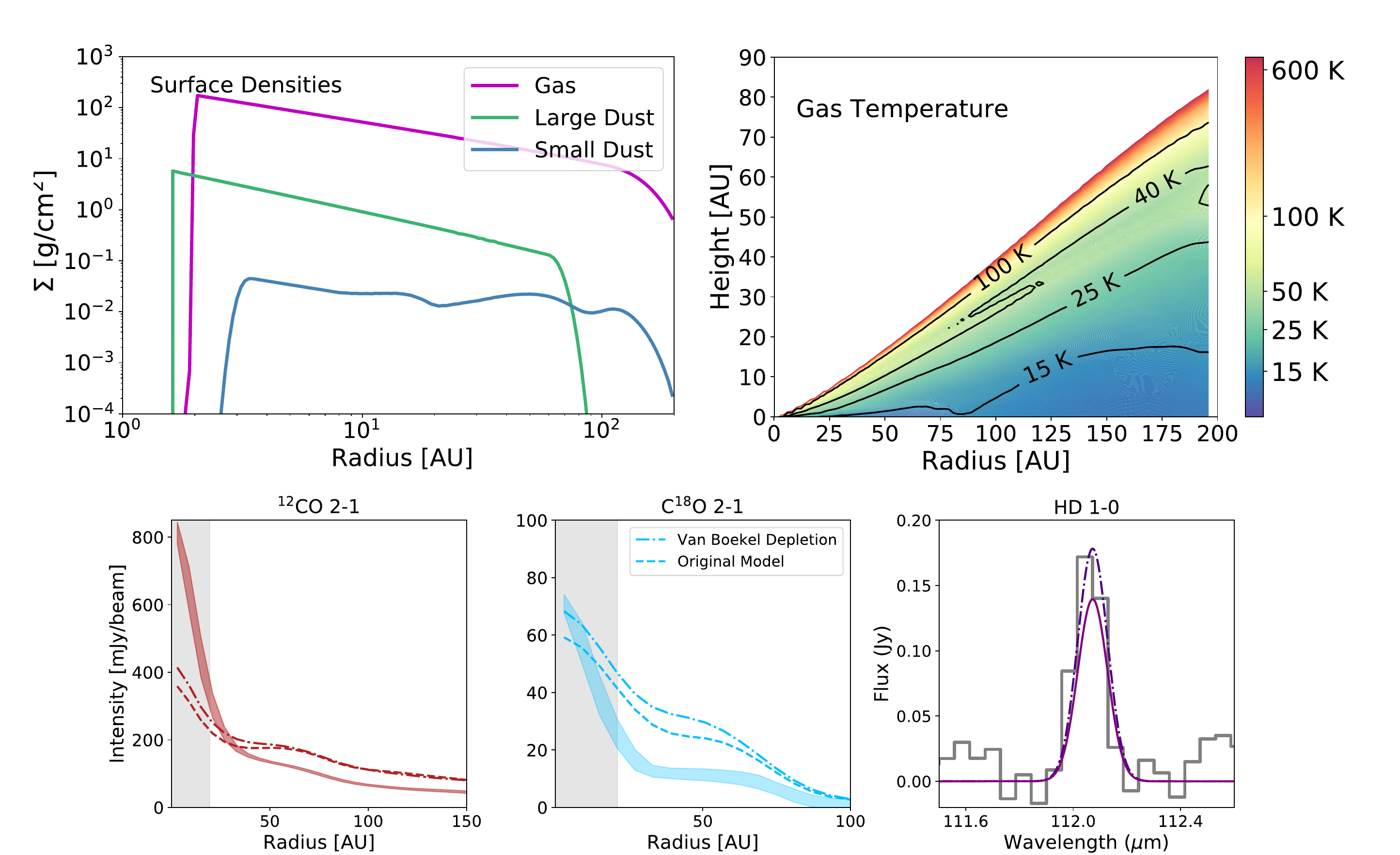}
    \caption{At the upper left are the surface densities for the gas and dust populations while taking a depletion of small dust as measured by \citet{vanBoekel17} from scattered light observations. Using the new small dust population, we calculate the gas temperature (top right), the radial profiles of \CO{} and \CetO{} 2-1 (bottom middle, left), and HD flux. The dashed lines in the radial profile and HD flux plot show our initial model, compared to a model with a non-smooth small dust surface density as motivated by the scattered light observations. }
    \label{fig:VBdata}
\end{figure*}

TW Hya has been observed in scattered light, which traces out the small dust population of the disk. These observations are reported in \citet{vanBoekel17}, and using the scattered light observations they find that radial depressions in the small dust surface density are necessary to reproduce the observed surface brightness. The two largest depressions in the surface density have depths 80\% and 45\% at approximately 5 AU and 90 AU, respectively  \citep[see Figure 11 in][]{vanBoekel17}. We applied this depletion factor to the small dust population to the initial model and found that it did not strongly effect the radial profiles as shown in Figure \ref{fig:VBdata}. The depletion does increase the HD $J$ = 1-0 flux by 30\%.  However, the effect on $^{12}$CO and C$^{18}$O is much less (below 15\%), and these changes do not significantly alter the fit to the CO radial profiles. Future modeling efforts may want to include information from the mm-continuum and scattered light observations as an additional constraint on the physical state of surface layers.  

\subsection{Comparison to Other Models}


There are at least two other attempts to model TW Hydra with thermo-chemical codes: \cite{Kama16_TWHya} with the DALI 2D physical-chemical code, and \cite{Woitke19} with the ProDiMo code.

\begin{figure}
    \centering
    \includegraphics[scale=.50]{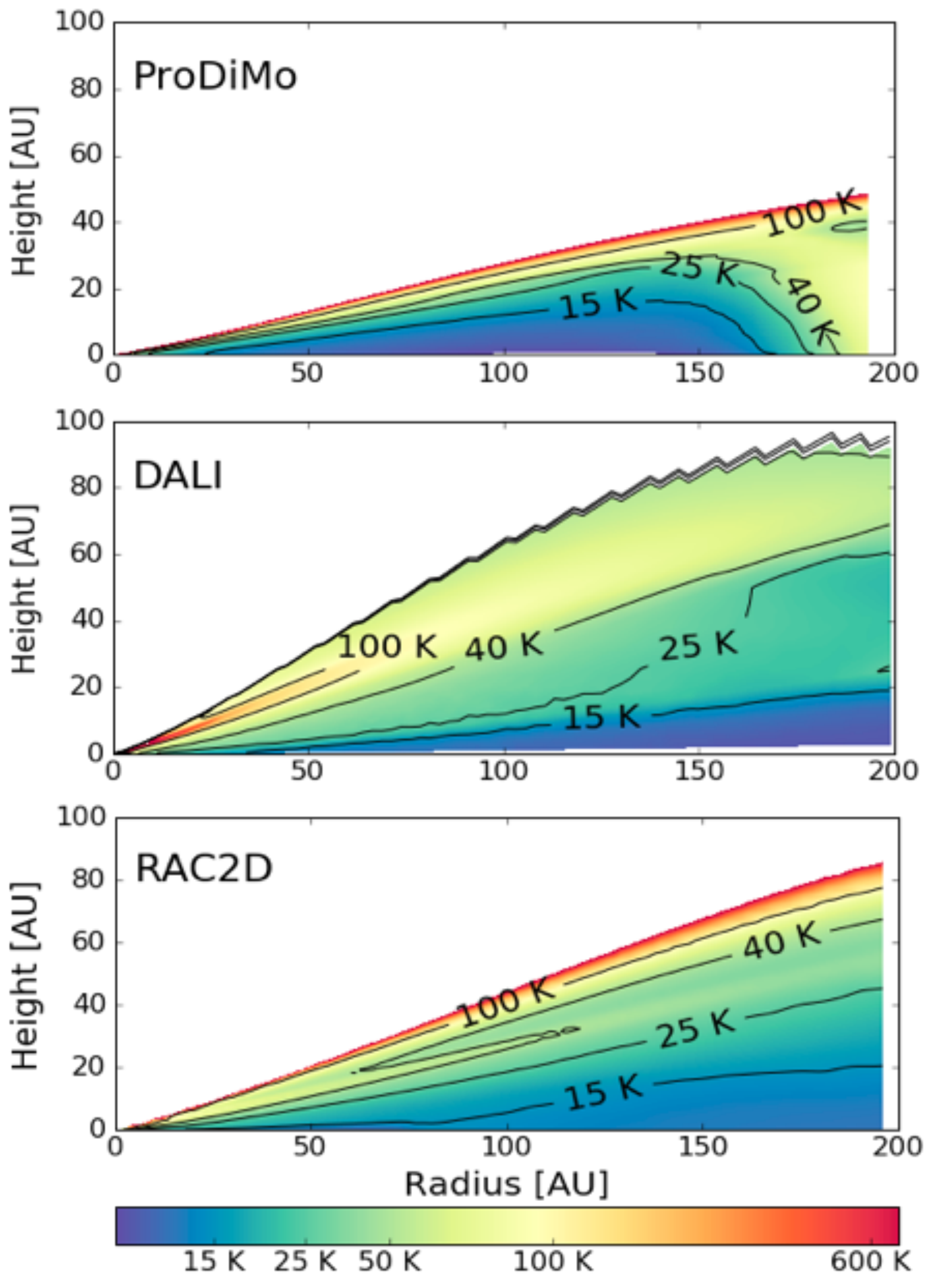}
    \caption{A comparison of the gas thermal structure of TW Hya from \citet{Woitke19}, \cite{Kama16_TWHya}, and this study.}
    \label{Compare}
\end{figure}

\begin{figure}
    \centering
    \includegraphics[scale=.45]{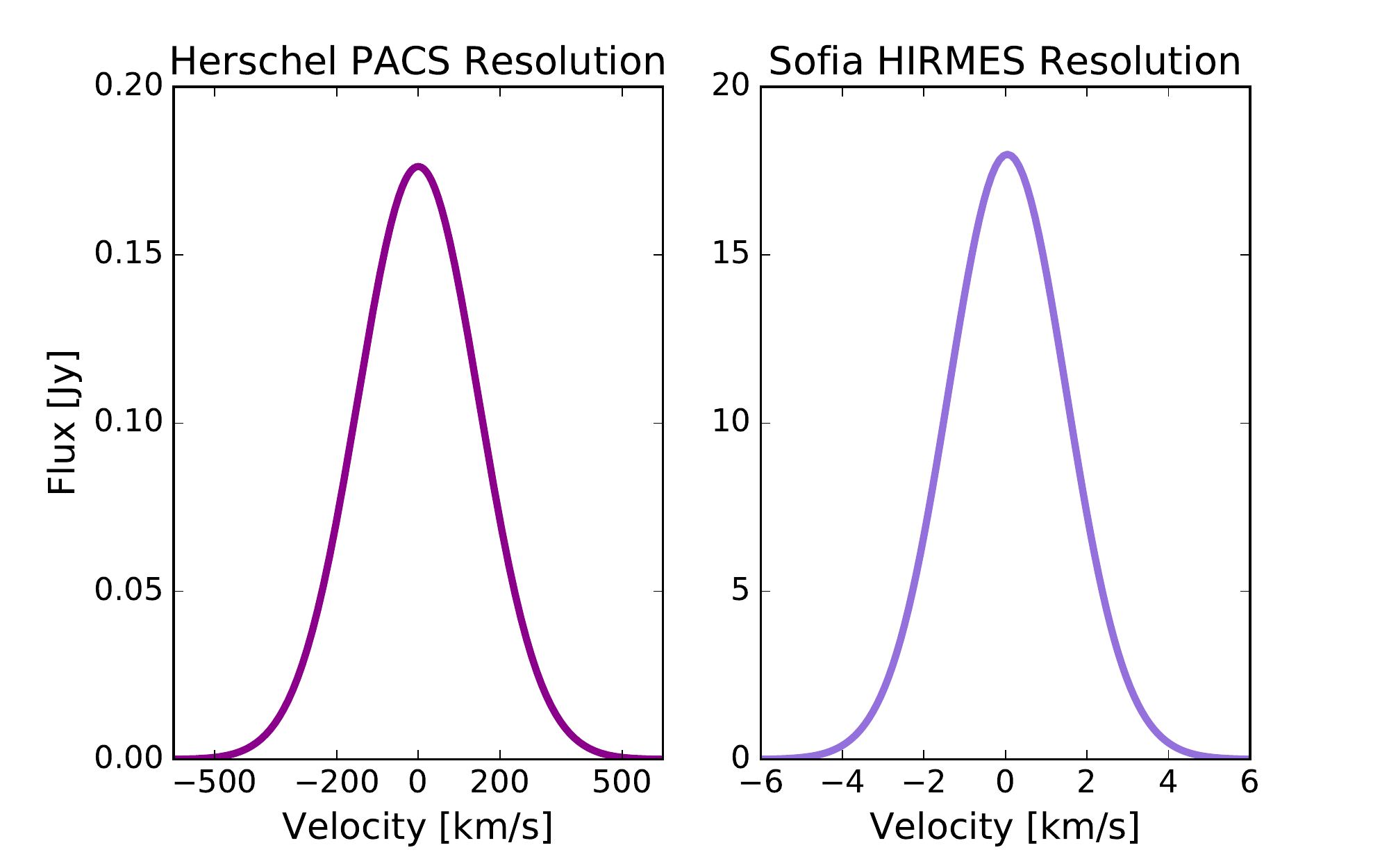}
    \caption{A predicted HD spectra based on the resolution of the proposed HIRMES instrument on \textit{Sofia} (right) compared to the modeled spectra of HD \textit{J}=1-0 from this study using the spectral resolution of the PACS instrument.}
    \label{prediction}
\end{figure}

\cite{Kama16_TWHya} focused on tracing volatile carbon in TW Hya. To do so, they modeled an underlying thermal structure of both the gas and dust population using DALI \citep{Bruderer09,Bruderer12,Bruderer13} which is designed and operated in a very similar way to RAC2D. This study utilized spatially unresolved observations of the CO ladder (upper limits of $J$=36-35, 33-32, 29-28, and line fluxes of $J$=18-17, 10-9, 6-5, ALMA observation of $J$=3-2, and SMA observation of $J$=2-1). Their observation of CO $J$ 3-2 was spatially resolved. Additionally, they utilized other carbon carrier molecules (C[I], C[II], \(\textrm{C}_{2}\textrm{H}\), \(\textrm{HCO}^{+}\)) to motivate their models along with HD both J=1-0 and J=2-1. \cite{Kama16_TWHya} produced a model that reproduces their CO observations and the HD flux. DALI has since been updated to include deuterium chemistry and isotoplogue selective chemistry following work by \citet{Miotello14}. This subsequent updated model is presented in \citet{Trapman17} and shown in Figure \ref{Compare}, alongside our RAC2D results. The main difference between \citet{Kama16} model and ours is that we reproduce multiple spatially resolved CO transitions. Our \(\sim\)15 - 100~K isotherms match up relatively well, which are the temperature ranges that our CO lines trace. Our solutions do diverge from each other at high temperatures (\(>100 \)K).  Using the HD emission, \cite{Kama16_TWHya} derived gas mass, and determined the same value that this study independently arrived at: 0.023 \(\textrm{M}_{\odot}\). 

\cite{Woitke19} describes the DIANA project, which modeled a number of protoplanetary disks in a homogeneous way using SED-matching and the thermo-chemical code ProDiMo. Their model outputs for each disk are publicly available. Figure \ref{Compare} shows a comparison of their results with our RAC2D model. The most noticeable difference is that our model appears to have a higher gas scale height compared to that determined by the DIANA project.  Further, the DIANA model finds higher gas temperatures at the disk surface, along with a comparatively cooler midplane (e.g. compare 15 K isotherms). \cite{Woitke19}  matched many volatile lines such as CO and \(\textrm{H}_{2}\textrm{O}\) to within a factor of two. However they under predicted the HD emission by a factor of 13 with a total gas mass of 0.045 \Msun{}. 
 
\citet{Cleeves15} also produced a TW Hya model with the primary goal of fitting molecular ion observations. In the process, this paper used SMA observations of \CetO{} and \thCO{} J=2-1 and the Herschel HD observation to constrain the physical structure. The chemical model employed was not a self consistent thermo-chemical model, however, the gas temperature structure was derived via a DALI-based fitting function to account for the gas and dust thermal decoupling in the upper layers.  In the \citet{Cleeves15} model, the derived gas mass of 0.04$\pm$0.02\,M$_\odot$ is also in good agreement with our best fit model, as is the overall temperature profile.

\subsection{Future Observations of Spatially Resolved CO Line Emission in Inclined Disks}

This study serves as a proof of concept that observation and subsequent modeling of resolved CO lines and HD line flux in combination with the SED is a powerful method to uncover thermal structure and disk mass. TW Hya is an ideal disk to test out this new method, as it is has been extensively observed with ALMA and other observatories, and is one of only three disks with an HD emission detection \citep{McClure16}. One disk parameter not explored in this particular study is disk inclination. An inclined disk will allow for direct constraints to be set on the vertical emitting layer and the subsequent temperature inferred within the layer \citep{Pinte18, Teague19}.  Disk inclination will broaden the spectral line, and the degree at which that might complicate extraction of a 2D thermal structure using thermo-chemical models is worth investigation. It is also worth applying these techniques towards brighter Herbig Ae/Be disks.  The CO snowline in these systems lies at greater distances from the star \citep{Qi15,Zhang20_hd163} and hence some transitions might probe layers closer to the midplane than sampled in T Tauri systems.

In our model we do not take CO fractionation chemistry into account. However, this would not effect the derived temperature structure very strongly. \CO{} would be unaffected and selective photodissociation has a marginal effect for \thCO{}.  There would be a larger effect for \CetO{}; however its emission is optically thin for much of the disk and thus does not provide sufficient temperature information. We also did not include rarer isotopologues of CO such as $^{13}$C$^{18}$O \citep{Zhang17}, which would provide additional insight into column density inside the CO snowline.

\subsection{Future HD observations}

At present, the sample of disks in which HD has been targeted is limited to a few deep observations \citep{McClure16} and those with full {\em Herschel} PACS scans, such as the upper limits on HD emission in Herbig disks by \citet{Kama20}. More observations are necessary to further our understanding of protoplanetary disks. There are future instruments and observatories potentially on the horizon that will observe the frequency range in which HD resides, and with a much higher resolution and sensitivity than \textit{Herschel}. These observatories will have the ability to drastically improve our understanding of early planet formation environments. The HIRMES instrument was proposed to be added to the SOFIA observatory offering a spectral resolution of 3 km s$^{-1}$ at 112 \(\micron\).  If built and deployed, HIRMES was projected to have a sensitivity similar to PACS \citep{Richards19} and could perform deep surveys for HD emission, opening up the ability to provide a robust diagnostic of gas mass. The planned spectral resolution on HIRMES is enough to fully resolve the HD emission line for inclined disks, which will provide unique constraints on the radial distribution of disk gas. A simulated HIRMES observation of TW Hya based on this model is shown in Figure \ref{prediction}.  

Looking onward, National Aeronautics and Space Administration's (NASA) \textit{Origins Space Telescope}, if funded and launched, provides a sensitivity improvement at 112 \micron{} of over three orders of magnitude compared to \textit{Herschel}, enabling deep surveys of over a thousand disks. Origins could provide sensitivities towards the HD line corresponding to gas disk masses down to \(10^{-5}\) \Msun \citep{Trapman17}. These future or potential observatories in combination with future resolved suites of CO lines via ALMA will be incredibly insightful towards out understanding of  planet formation enabling the ability to track the 2D temperature structure and the gas mass, thereby revealing the key physical conditions under which planets form.

\section{Conclusions} \label{sec:Conclusions}
Using the thermo-chemical code \textit{RAC2D}, and observations from the TW Hya Rosetta Project taken with ALMA of \CO{} $J$= 2-1, \CetO{} $J$= 2-1, archival ALMA data of \CO{} 3-2, \CetO{} 3-2, 6-5, and \thCO{} 3-2, 6-5, HD $J$ 1-0, and the system's SED, we find the following:

\begin{enumerate}
    \item We derive a thermal structure of the TW Hya disk that fits radial brightness profiles of seven spatially resolved CO line transitions, a spatially unresolved HD line flux, and the SED. This study is the first to reproduce multiple spatially resolved observations of CO using a thermo-chemical model. We compare our results to thermal structures derived by other methods.
    \item Using the derived thermal structure and the HD (1-0) line flux, we constrain the total gas mass of the TW Hya disk to be \(\approx\) 0.025 \Msun{} within our model framework.
    \item We directly demonstrate that CO line brightness distributions are degenerate results of the CO abundance and the total gas mass of the disk. Therefore CO alone cannot be used as a robust tracer of the total gas mass, for any protoplanetary disk system. We find that using observations of HD is one solution that breaks this degeneracy.
\end{enumerate}

\section{Acknowledgments}
This paper makes use of data from ALMA programs 2015.1.00686.S, 2016.1.00629.S, 2012.1.00422.s, and 2016.1.00311.S. ALMA is a partnership of ESO (representing its member states), NSF (USA) and NINS (Japan), together with NRC (Canada) and NSC and ASIAA (Taiwan), in cooperation with the Republic of Chile. The Joint ALMA Observatory is operated by ESO, AUI/NRAO and NAOJ. The National Radio Astronomy Observatory is a facility of the National Science Foundation operated under cooperative agreement by Associated Universities, Inc.
J.K.C. acknowledges support from the National Science Foundation Graduate Research Fellowship under Grant No. DGE 1256260 and the National Aeronautics and Space Administration FINESST grant, under Grant no. 80NSSC19K1534. 
E.A.B. acknowledges support from NSF grant No. 1907653.
K.S. acknowledges the support of NASA through Hubble Fellowship grant HST -HF2-51401.001 awarded by the Space Telescope Science Institute, which is operated by the Association of Universities for Research in Astronomy, Inc., for NASA, under contract NAS5-26555. 
L.I.C. gratefully acknowledges support from David and Lucile Packard Foundation and the Virginia Space Grant Consortium.
J. H. acknowledges support from the National Science Foundation Graduate Research Fellowship under Grant No. DGE-1144152. S. A. and from the National Aeronautics and Space Administration under Grant No. 17-XRP17 2-0012 issued through the Exoplanets Research Program. Support for this work was provided by NASA through the NASA Hubble Fellowship grant \#HST-HF2-51460.001-A awarded by the Space Telescope Science Institute, which is operated by the Association of Universities for Research in Astronomy, Inc., for NASA, under contract
NAS5-26555.
C.W. acknowledges financial support from the University of Leeds and the Science and Technology Facilities Council (STFC; grant No. ST/R000549/1).
M.K. gratefully acknowledges funding by the University of Tartu ASTRA project 2014-2020.4.01.16-0029 KOMEET ``Benefits for Estonian Society from Space Research and Application'', financed by the EU European Regional Development Fund

\facilities{ALMA, Herschel}
\software{RAC2D \citep{Du14}, bettermoments\citep{Teague18}, CASA \citep{McMullin07}}


\appendix
\restartappendixnumbering

\section{Parameter Effects on Simulated Observations}\label{sec:effects}

Through exploration of our model and the parameter space of the properties listed in Table \ref{paramsrange}, we find that CO emission is most strongly affected by gas and small dust parameters. Here we list how the CO integrated intensity profiles respond to changes in parameters.  These findings are specific to this TW Hya model, but can be extrapolated to similarly inclined gaseous disks. The following discussion is based off of the model parameter values from \citet{Zhang19} presented in Figure \ref{fig:original}: 
\begin{figure}[h]
    \centering
    \includegraphics[scale=.4]{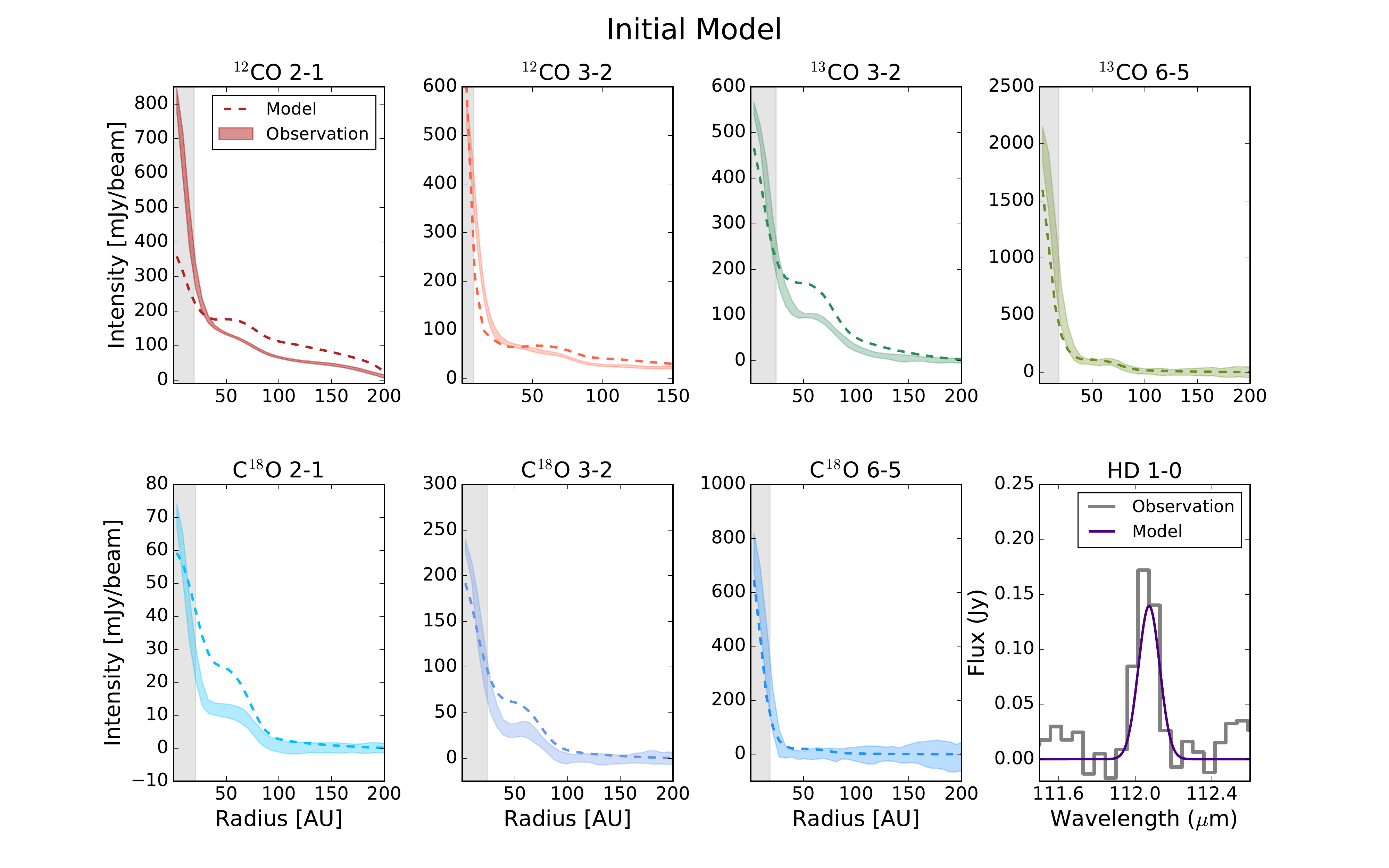}
    \caption{The integrated line profiles of CO and its isotopologues from ALMA observations (solid lines) and the simulated observations from RAC2D using model parameters from \citet{Zhang19}, and listed in Table \ref{paramsrange}}
    \label{fig:original}
\end{figure}

\subsection{$\gamma$ : Power-law Index of Surface Density}
\(\gamma\) is a power-law index for the surface density (see Equation \ref{surfden}) with a maximum value of two. Increasing \(\gamma\) affects the distribution by concentrating the given component (gas, dust) towards the inner disk region, increasing the column density of gas and dust. A decrease in \(\gamma\) results in a more even distribution of the mass component. It has the strongest effect on the emission arising from the inner 25 AU, especially for the rare isotopologues. Altering gamma from 0.75 to 0.9 in both small dust and gas results in very little change in \CO{} 2-1 and 3-2, but at least a 10\% increase across all \CetO{} lines. In our final model, we increased \(\gamma\) in both the gas and small dust, as they are coupled, from 0.75 to 0.85 which aided in adding to the intensity in the inner 25 AU. 

\begin{figure}
    \centering
    \includegraphics[scale=.4]{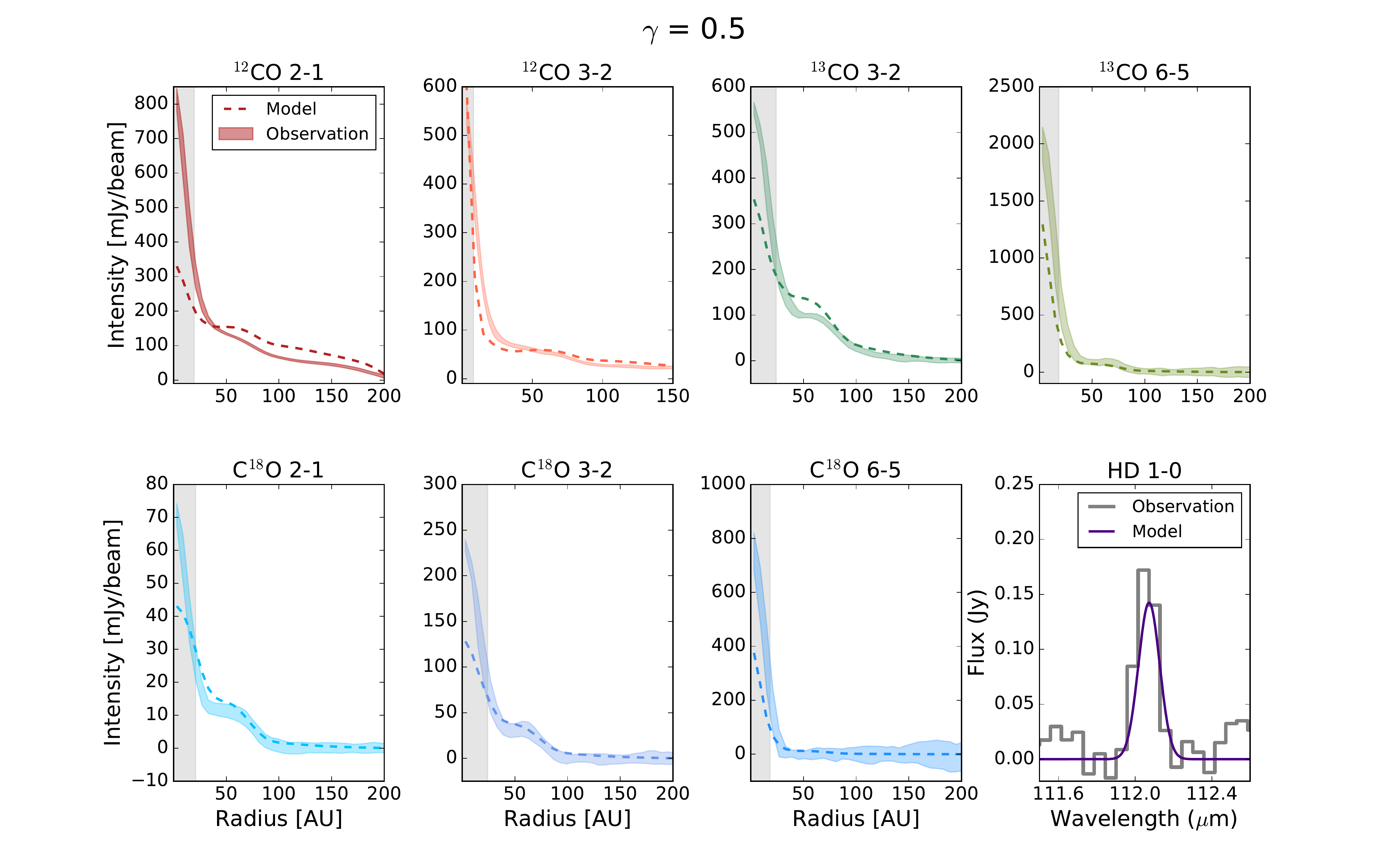}
    \caption{The integrated intensity profiles of observations and a model based of the initial model (\ref{fig:original}) but with $\gamma$ in the gas and small dust = 0.5. }
    \label{fig:gam0.5}
\end{figure}

\begin{figure}
    \centering
    \includegraphics[scale=.4]{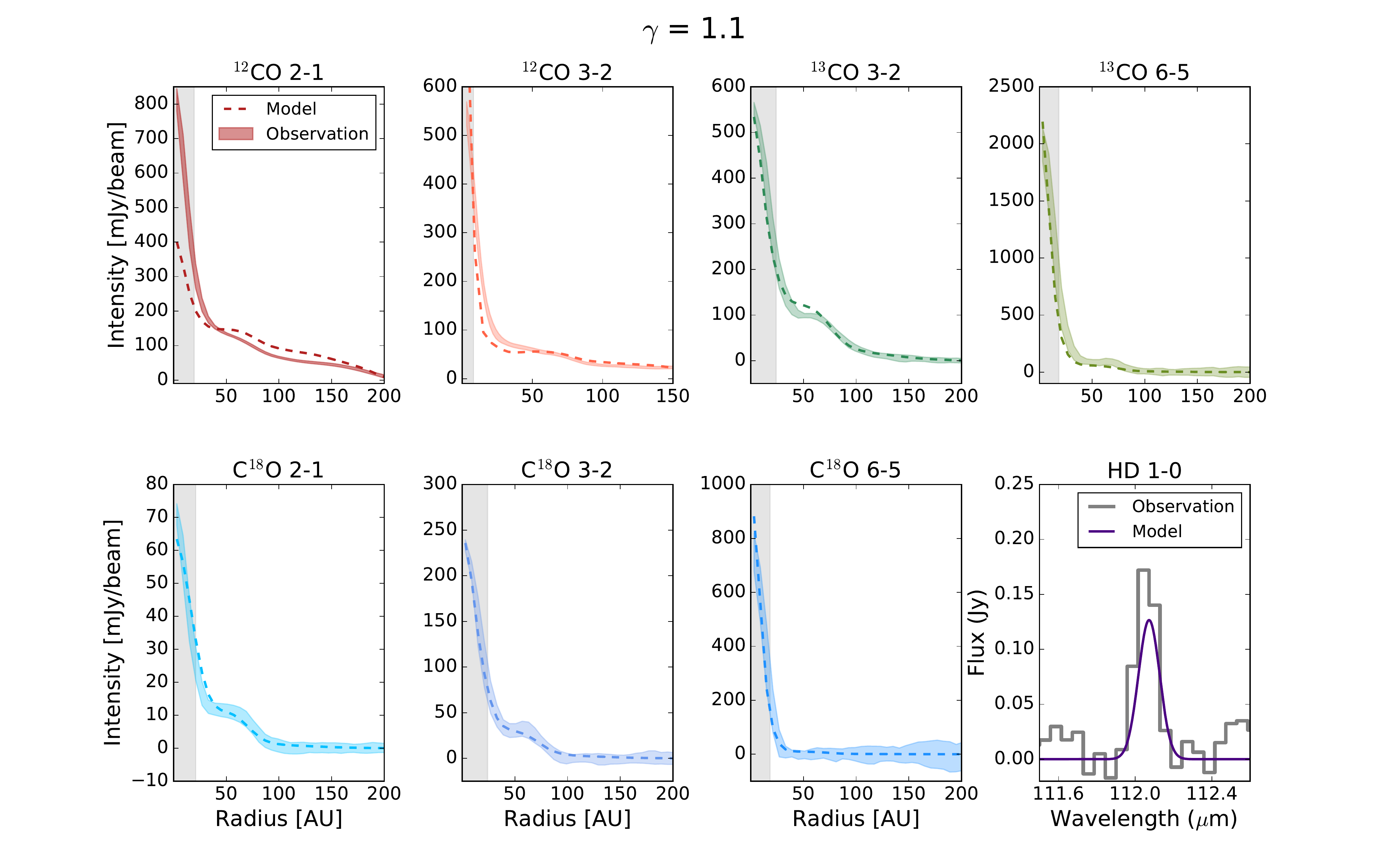}
    \caption{The integrated intensity profiles of observations and a model based of the initial model (\ref{fig:original}) but with $\gamma$ in the gas and small dust = 1.1. }
    \label{fig:my_label}
\end{figure}

\subsection{ $\Psi$ : Power-law Index For Scale Height}

Typical \(\Psi\) values are between 1.1 and 1.2 for TW Hya, and affects the scale height over different radii, with an increase in the flaring as \(\Psi\) increases. Changes in \(\Psi\) down to 0.05 have a significant effect on the final integrated intensity profiles. Generally, lowering \(\Psi\) results in an increase in intensity in the inner disk (\(<\) 25 AU), and beyond \(\approx\) 25 AU features are `smoothed' ouft. This is due to the increase of gas surface density towards the inner AU when flaring decreases. The most extreme value of \(\Psi\) we explored was 1.6, and at that point the modeled intensity profiles plummeted down to between 25\% of their original flux to zero within 20 AU. Beyond 20 AU, there is emission comparable to the original model. The thermal profile derived from the initial parameters (shown in Table \ref{paramsrange} and discussed in Section \ref{depletion}) produced CO lines that were too dim within 25 AU by a factor of 2 in \CO, \thCO{} 6-5, \CetO{} 6-5 and \(\sim\) 30\% in \thCO{} and \CetO{} 3-2. We find that the only way to significantly brighten the inner regions while simultaneously keeping the intensity in the outer regions low is to allow the gas and small dust to have slightly different \(\Psi\) values. The gas component is given a \(\Psi\) value of 1.1, while the small dust is given \(\Psi\)= 1.2. Due to the fact that our critical radius is beyond the gaseous radius, even though the dust has a higher flaring angle, it lies below the gas. This creates a thin region at the upper layer of the disk where only gas resides; this region has a thickness of \(\sim\) 1.2\% of a given radius. In this gas-rich layer, CO and HD lines become brighter, noticeably within the inner few AU. There are no other parameters that we explored that accomplished this, the closest being \(\gamma\), which increases emission within 50 AU, but smooths out the feature beyond 50 AU. 

\begin{figure}
    \centering
    \includegraphics[scale=.4]{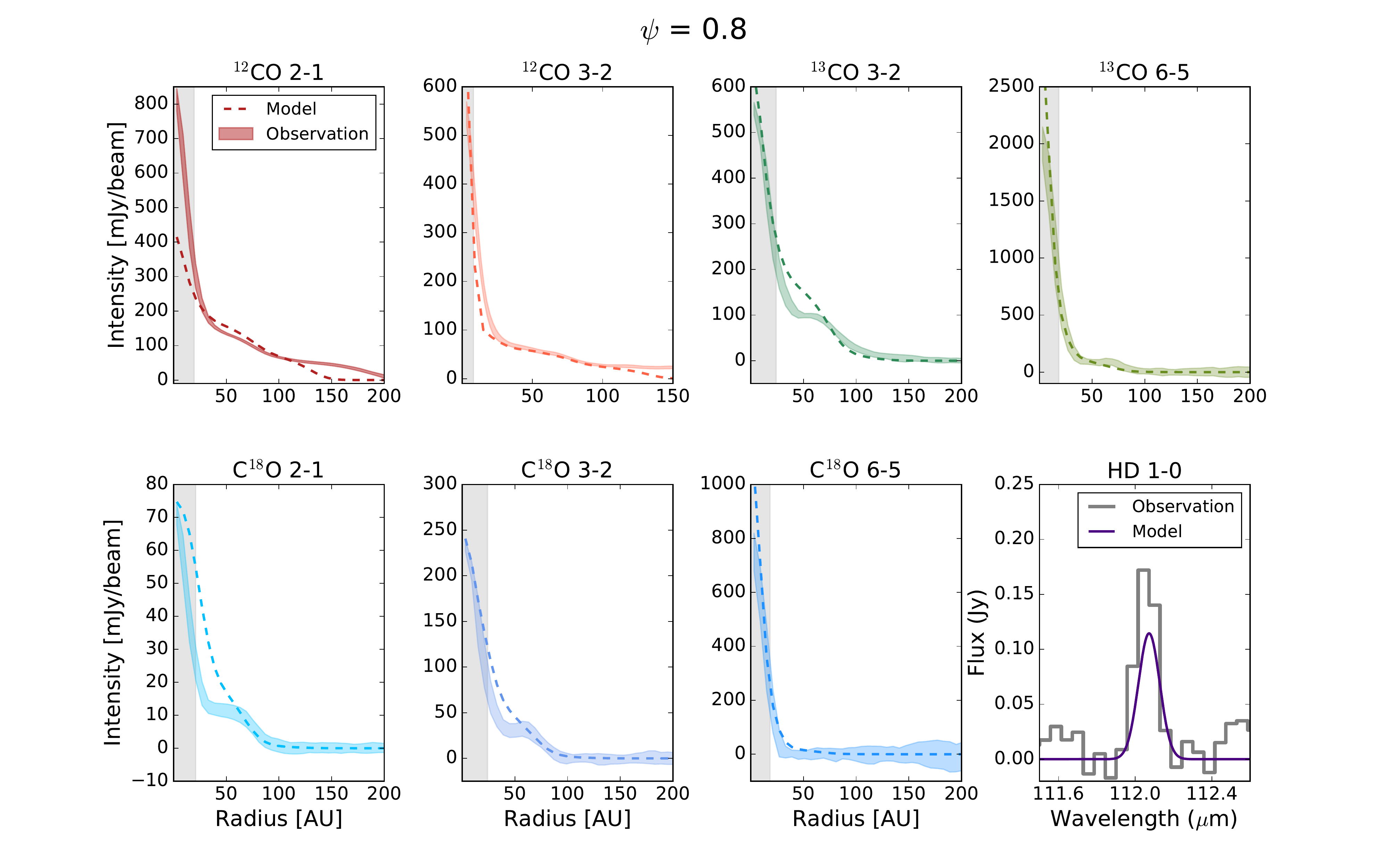}
    \caption{The integrated intensity profiles of observations and a model based of the initial model (\ref{fig:original}) but with $\Psi$ in the gas and small dust = 0.4. }
    \label{fig:psi_0.8}
\end{figure}

\begin{figure}
    \centering
    \includegraphics[scale=.4]{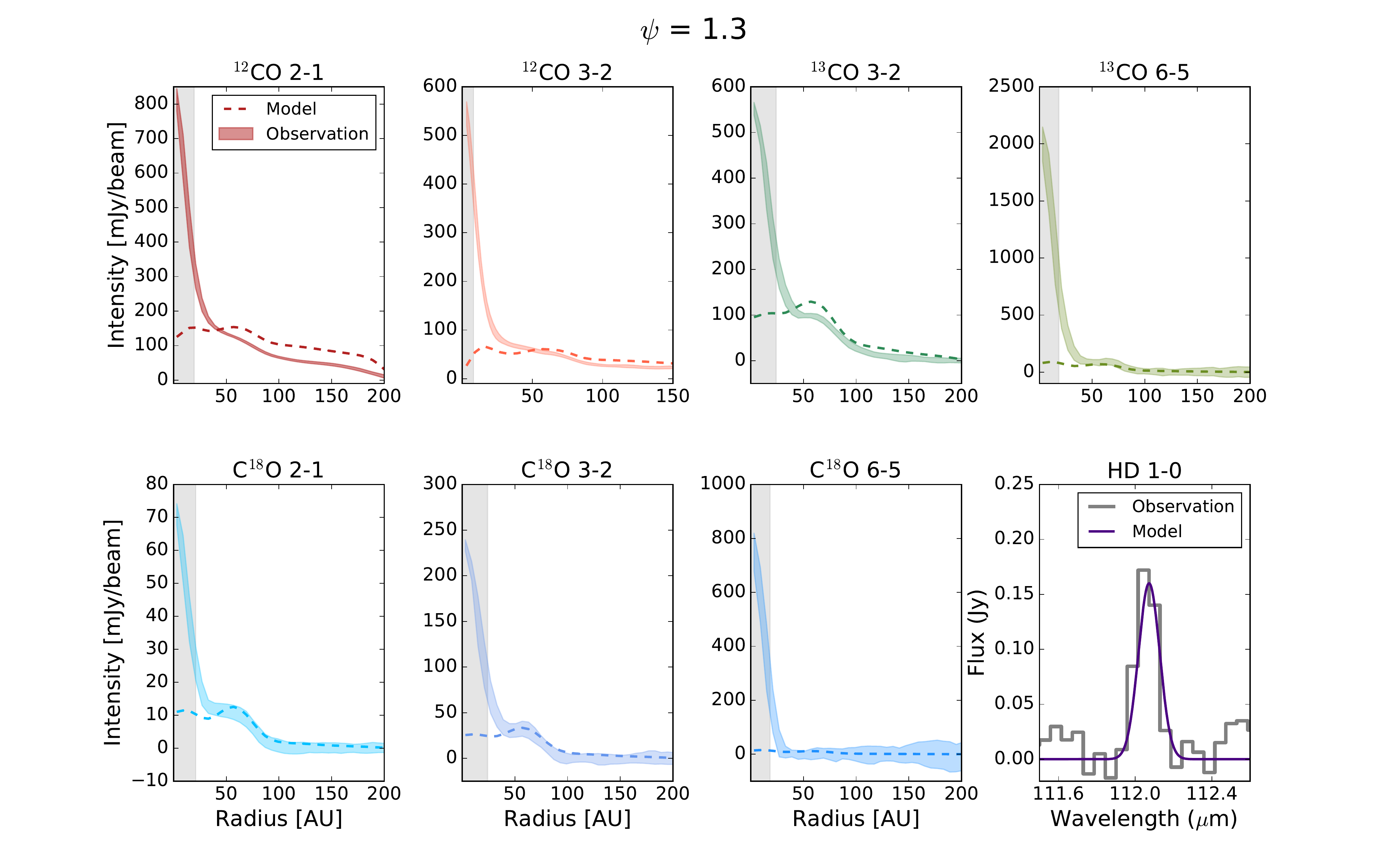}
    \caption{The integrated intensity profiles of observations and a model based of the initial model (\ref{fig:original}) but with $\Psi$ in the gas and small dust = 1.6. }
    \label{fig:psi_1.3}
\end{figure}

\subsection{Gas Mass}
An increase in the total gas mass increases the intensity from each line along all radii. A higher gas mass results in a higher column density for each molecule, and some of the CO lines (especially \CO{} 2-1) are already optically thick in explored  mass range.  There are also isotopologues whose emission is not fully optically thick in the lower mass models, thus changes in their column have a strong effect on their flux. Increasing the gas mass from 0.01 to 0.02 \(\textrm{M}_{\odot}\) doubles the peak integrated flux in \CetO{}, a 25\% increase in \thCO{}, and \CO{} 3-2 and an increase of \(\sim\) 20\% in \CO{} 2-1. Our best-fit model has a disk mass of 0.023 solar masses. We arrived at this value only after the gas and small dust components of our model were given different \(\Psi\) values. With a gas mass of 0.05 \(\textrm{M}_{\odot}\) and \(\Psi_{\rm gas}\) = 1.1 and \(\Psi_{\textrm{small dust}}\) = 1.2, the HD flux increases by a factor of two, and emission in CO at large radii also increases by a factor of 2-3 depending on the isotopologue. Due to this, a decrease in total gas mass is justified. Decreasing the mass to 0.023 \(\textrm{M}_{\odot}\) brings the HD flux back down to what \textit{Herschel} observed, and brings the CO emission back down to values that agree with ALMA observations. 

\begin{figure}
    \centering
    \includegraphics[scale=.4]{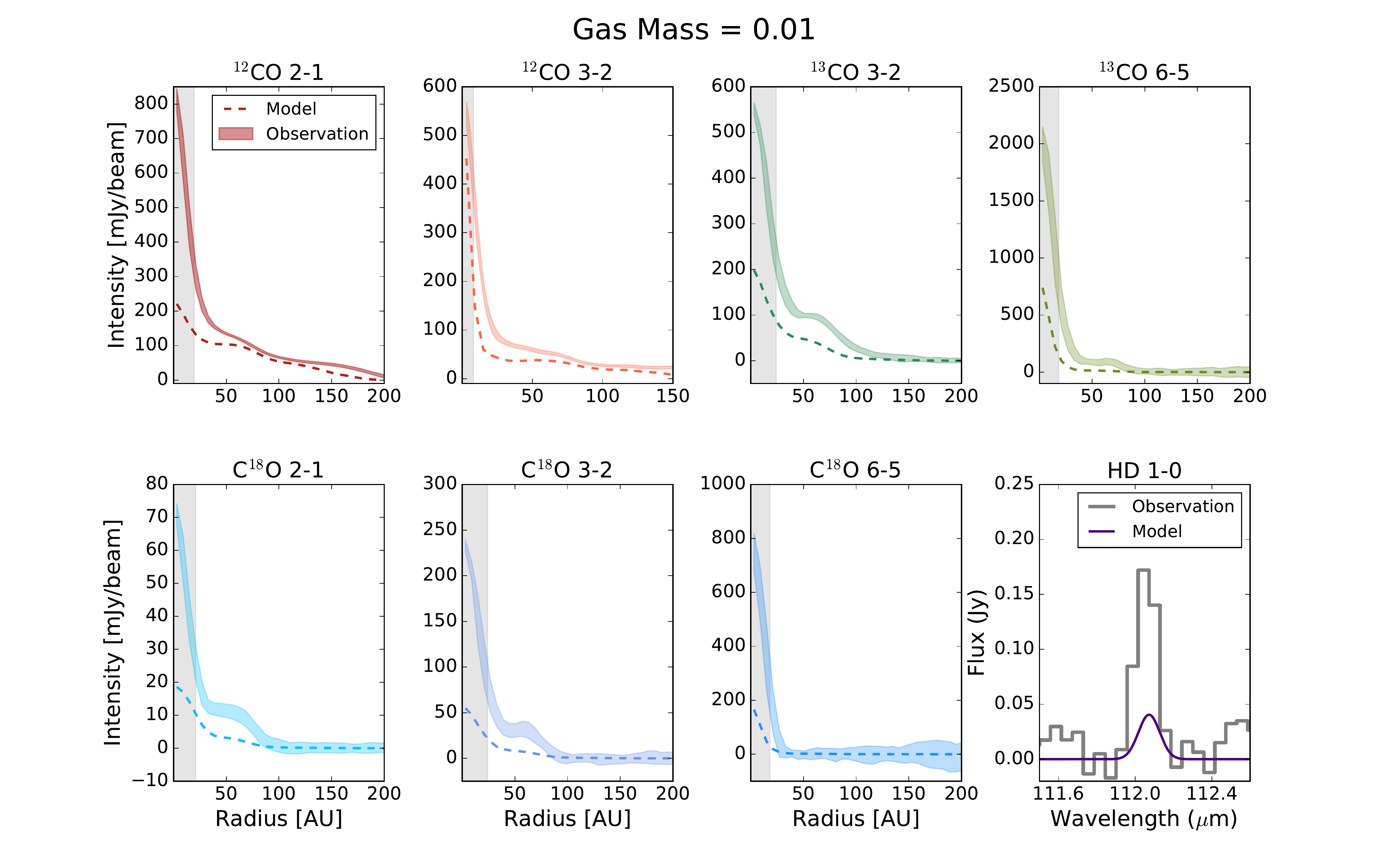}
    \caption{The integrated intensity profiles of observations and a model based of the initial model (\ref{fig:original}) but with a gas mass equal to 0.01 \Msun{} }
    \label{fig:gm_0.01}
\end{figure}

\begin{figure}
    \centering
    \includegraphics[scale=.4]{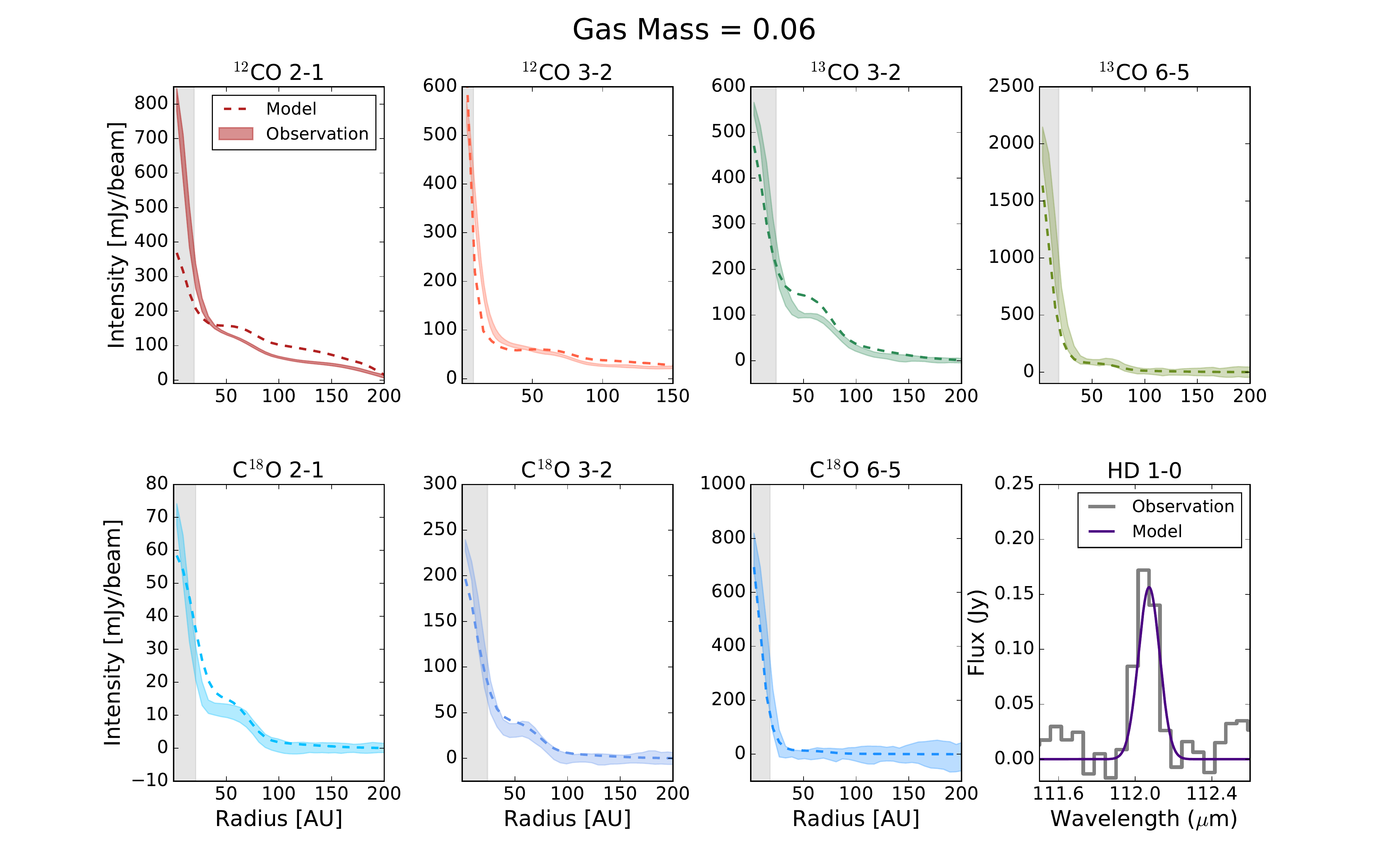}
    \caption{The integrated intensity profiles of observations and a model based of the initial model (\ref{fig:original}) but with a gas mass equal to 0.06 \Msun{}}
    \label{fig:gm_0.06}
\end{figure}

\subsection{Small Dust Component Mass}
Given a constant gas mass of 0.05 \(\textrm{M}_{\odot}\) and a starting small dust mass of 1.0 \(\times 10^{-4}\) \(\textrm{M}_{\odot}\) or higher, decreasing the small dust mass has a similar effect as increasing gas mass. At 1.0 \(\times 10^{-4}\) \(\textrm{M}_{\odot}\) and lower, flux beyond 20 AU increases at a faster rate than within 20 AU, leading to exaggerated features, such as the plateaus in \CetO{} 2-1 and 3-2 becoming peaks at small dust masses less than 2.5 \(\times 10^{-5}\) \Msun{}. The small dust population absorbs and scatters radiation  and in some instances leads to a CO line profile that is extincted. Eliminating small dust allows for the gas that exists to emit more freely as a major opacity source is removed. The small dust mass is particularly impactful on the HD emission, because the small dust governs the propagation of UV radiation, with smaller dust masses allowing UV photons from the star to penetrate deeper layers of the disk, directly affecting the gas heating terms. This helps populate the HD $J$=1 level, increasing the HD $J$=1-0 line flux. In our final model, we were not strongly motivated to change the small dust mass beyond what had been initially predicted, but it could be a useful lever in future modeling of disks.   At present the strongest constraint on the mass of the small dust is the SED and scattered light emission form the surface.

\begin{figure}
    \centering
    \includegraphics[scale=.4]{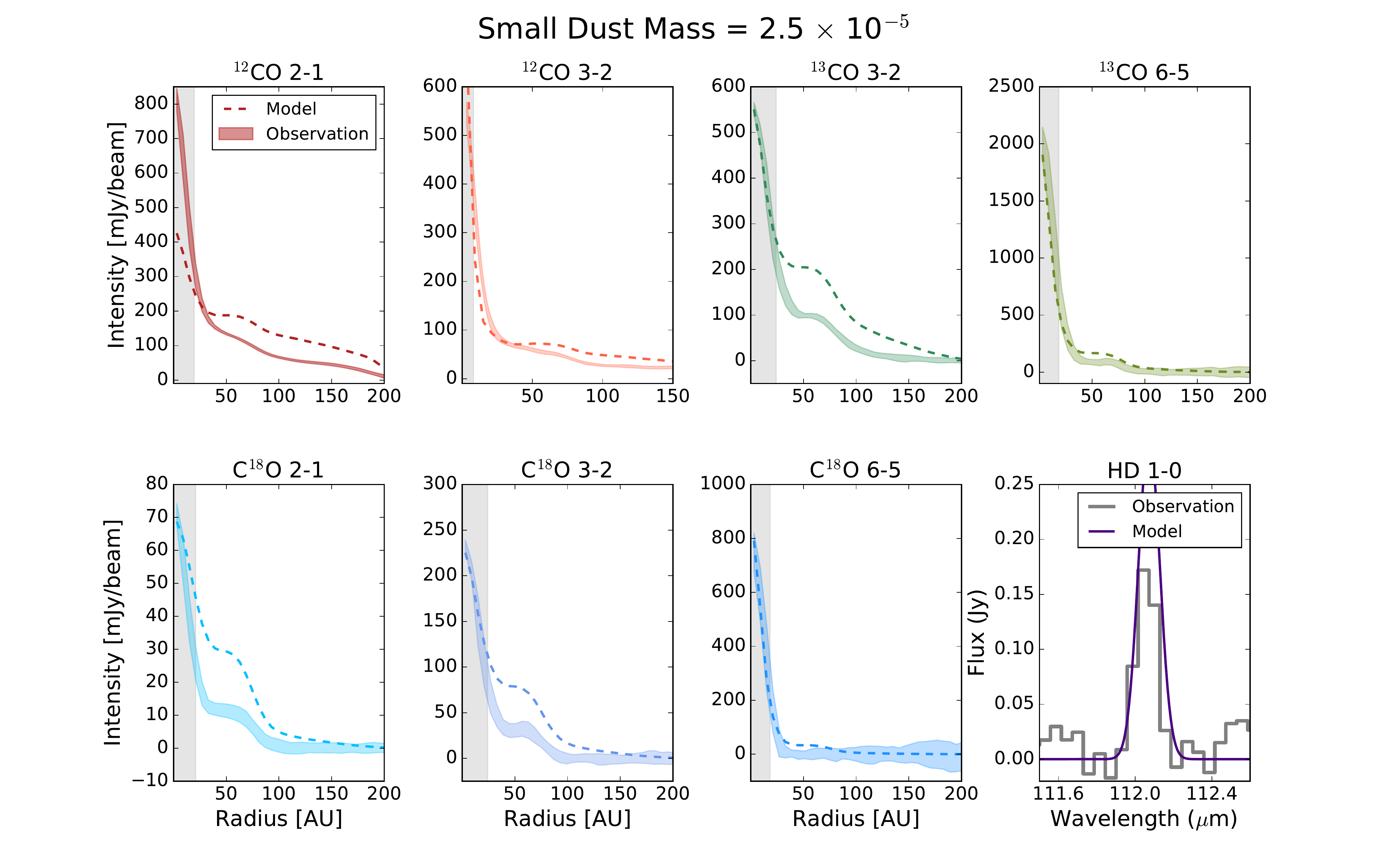}
    \caption{The integrated intensity profiles of observations and a model based of the initial model (\ref{fig:original}) but with a total small dust mass equal to 2.5 $\times$ 10$^{-5}$ \Msun{}}
    \label{fig:sd_2.5e-5}
\end{figure}

\begin{figure}
    \centering
    \includegraphics[scale=.4]{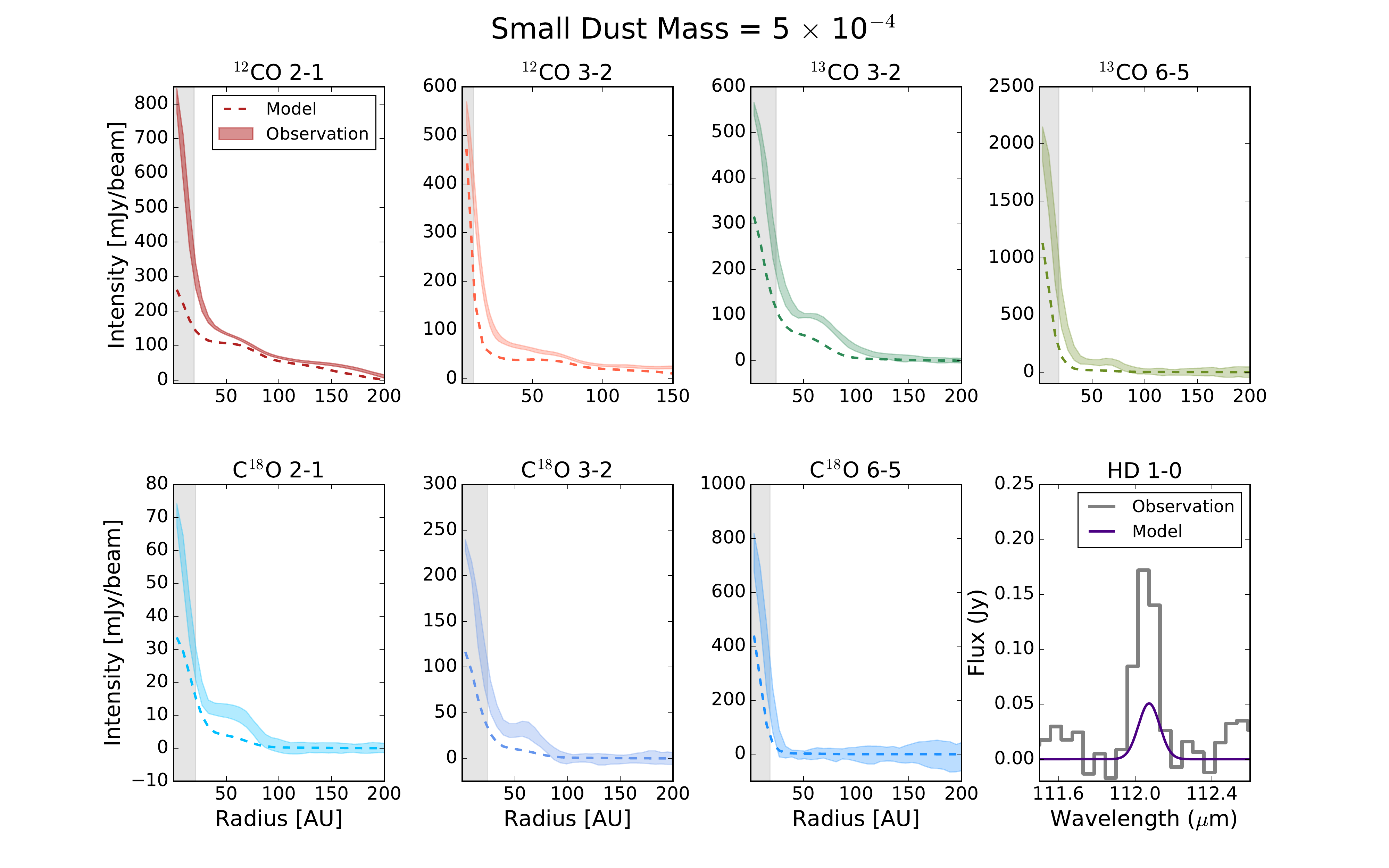}
    \caption{The integrated intensity profiles of observations and a model based of the initial model (\ref{fig:original}) but with a total small dust mass equal to 5 $\times$ 10$^{-4}$ \Msun{}}
    \label{fig:sd_5e-4}
\end{figure}

\subsection{\(\rm r_{\textrm{exp}}\) : Outer Tapering Radius}

At this specified radius the abundance of a component exponentially decreases. Altering the \(r_{\textrm{exp}}\) effects outer disk emission, as it significantly depletes the corresponding component. It is only significant in \CO{} 2-1 and 3-2, as these are the only intensity profiles with emission beyond 100 AU. If the gas and small dust \(r_{\textrm{exp}}\) value becomes less than the large dust \(r_{\textrm{exp}}\), the effect is much stronger and acts similar to decreasing \(\gamma\).

\begin{figure}
    \centering
    \includegraphics[scale=.4]{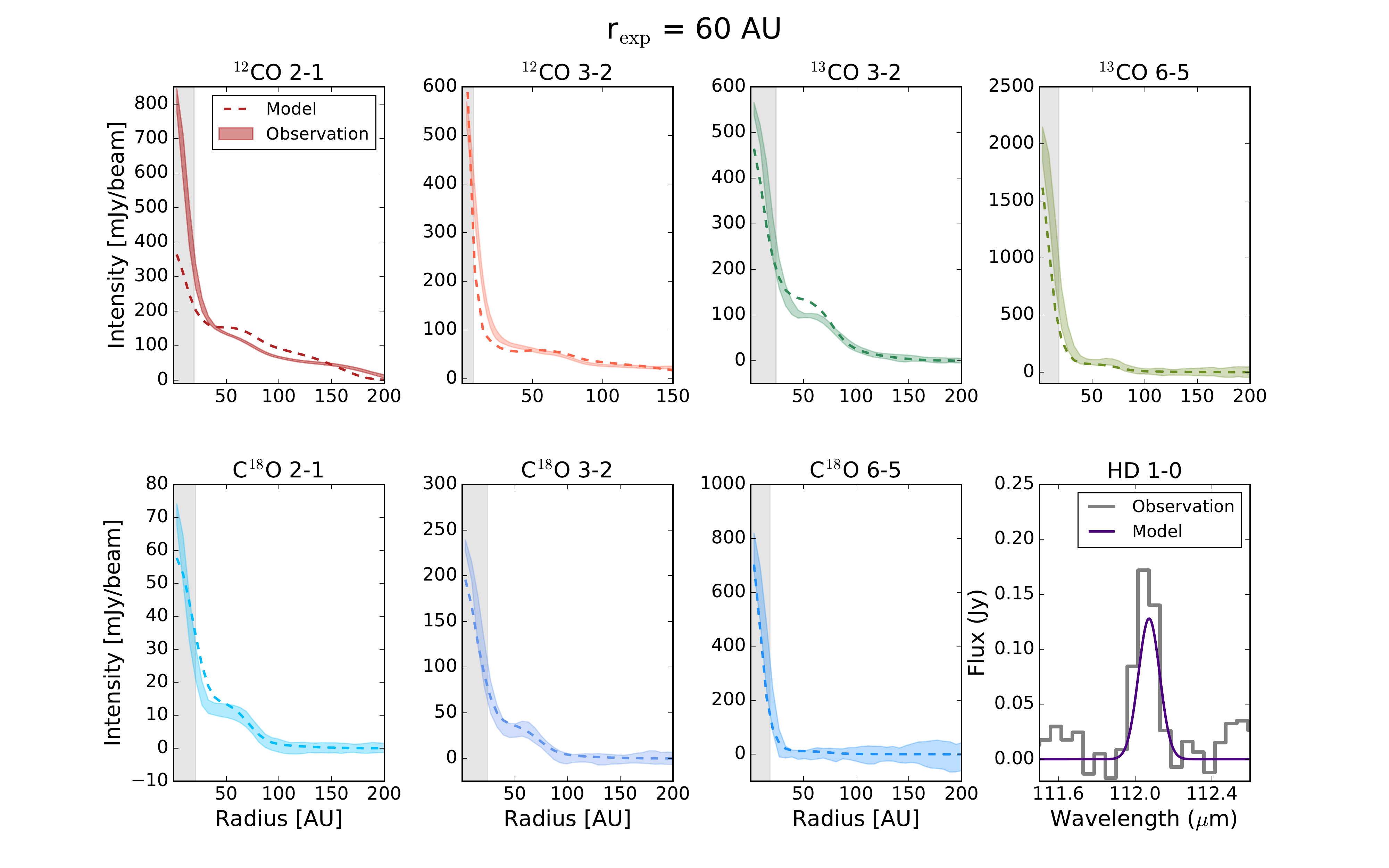}
    \caption{The integrated intensity profiles of observations and a model based of the initial model (\ref{fig:original}) but with r$_{\rm exp_{\rm out}}$ in the gas and small dust = 60 AU }
    \label{fig:rexp60}
\end{figure}

\begin{figure}
    \centering
    \includegraphics[scale=.4]{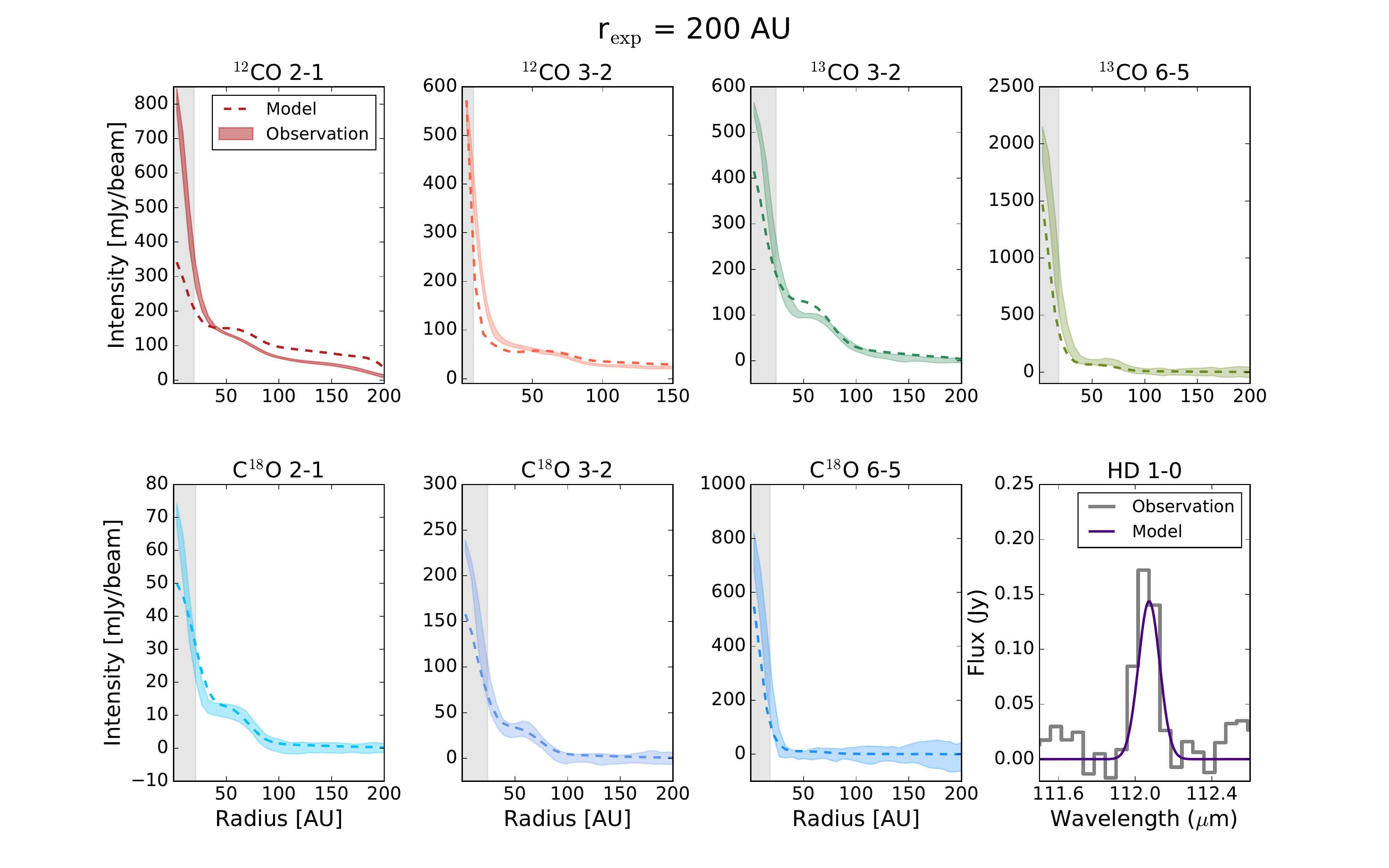}
    \caption{The integrated intensity profiles of observations and a model based of the initial model (\ref{fig:original}) but with r$_{\rm exp_{\rm out}}$ in the gas and small dust = 200 AU }
    \label{fig:rexp200}
\end{figure}

\subsection{$\rm h_{c}$: Characteristic Height}
An increase of the height of the disk results in an increase in intensity across all radii and in all molecules, with the ones most strongly affected being the least abundant. A change in scale height increases the column that one observes, which then increases the flux of an optically thin molecule. However, this does not strongly affect optically thick molecules. An increase of two times the scale height only results in a \(\sim\) 5 \% increase in peak intensity in \CO{} 2-1 and a full 20\% increase in the least common isotopologue transition considered here, \CetO{} 6-5.

The scale height, scale radius, \(\Psi\), and \(\gamma\) of the large dust population do not have strong effects on the CO line profiles. The temperature of the star also does not have a strong effect; a change in 1000 K resulted in only a ~1\% increase across all lines. 

The parameters we find having the largest impact on the CO and HD flux were: gas mass, the flaring parameter \(\Psi\), surface density distribution parameter \(\gamma\), and small dust mass. The gas mass affects all CO lines, and has a stronger influence on the optically thin lines. \(\Psi\) and \(\gamma\) both redistribute flux between the inner and outer regions of the disk depending on where the majority of the gas component is located. The small dust mass was not a parameter that needed to be altered from its initial value, but has a strong effect on the HD flux and could be utilized in future studies as long as the SED is taken into account.

\begin{figure}
    \centering
    \includegraphics[scale=.4]{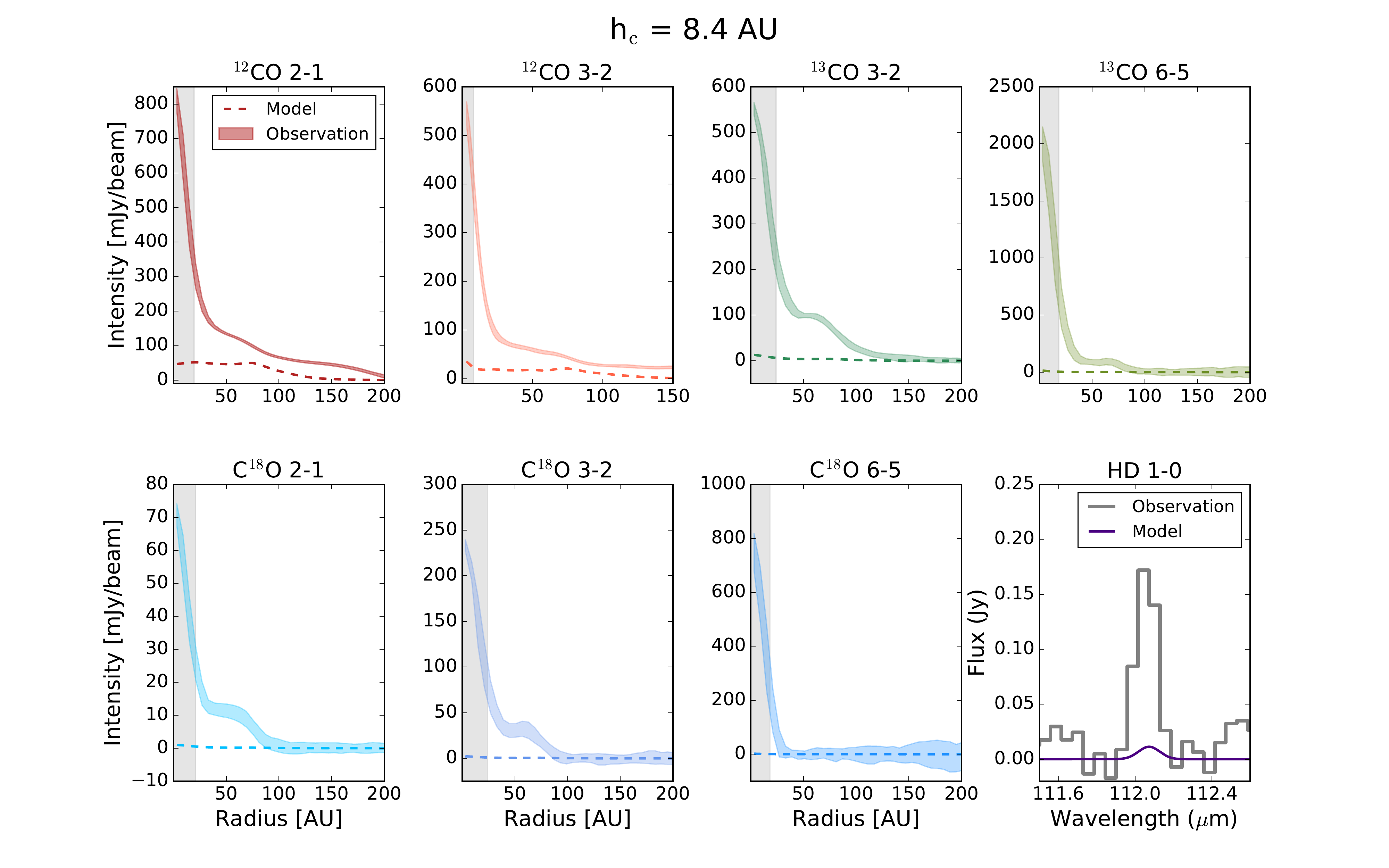}
    \caption{The integrated intensity profiles of observations and a model based of the initial model (\ref{fig:original}) but with h$_{c}$ in the gas and small dust = 8.4 AU }
    \label{fig:hc8.4}
\end{figure}

\begin{figure}
    \centering
    \includegraphics[scale=.4]{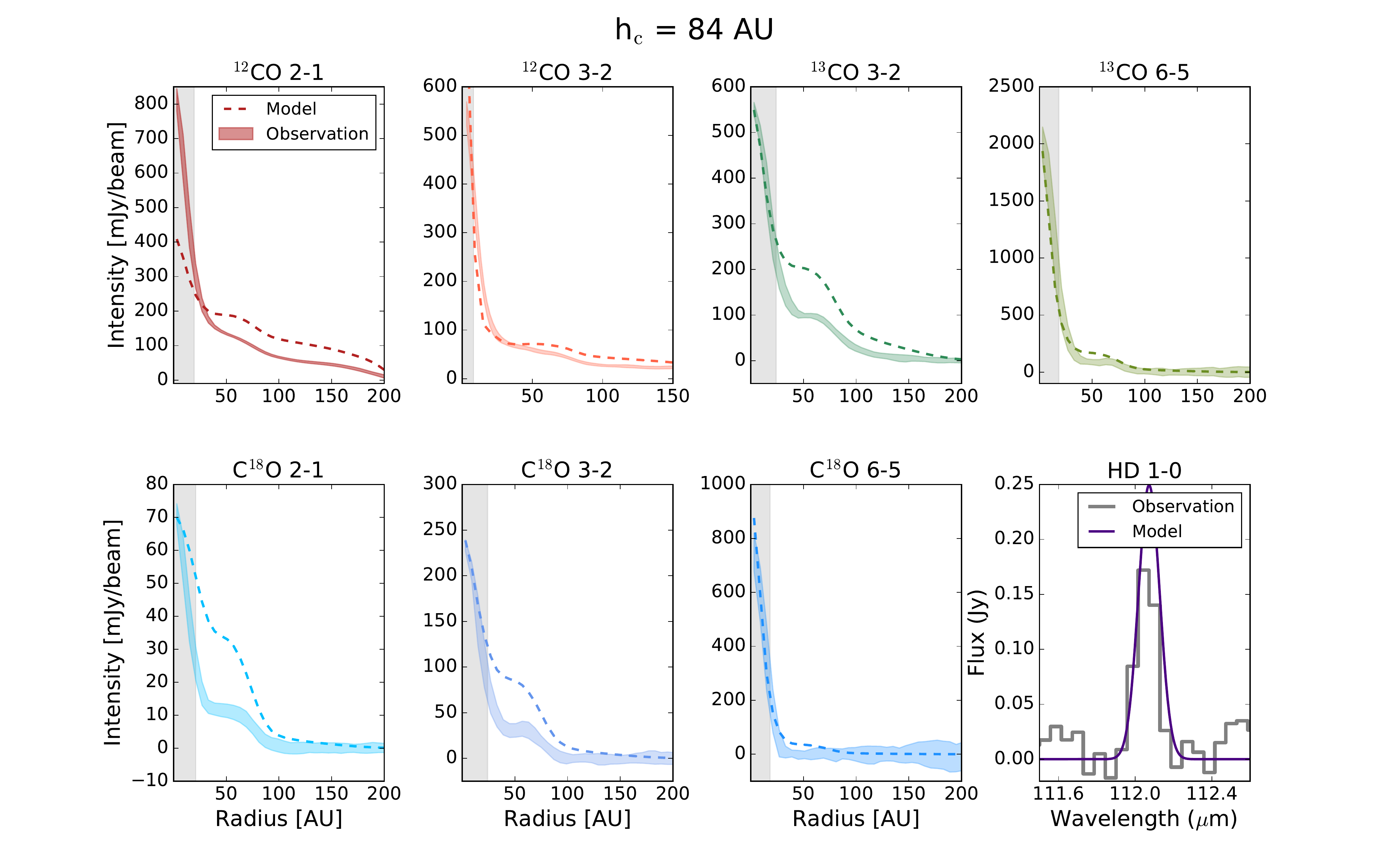}
    \caption{The integrated intensity profiles of observations and a model based of the initial model (\ref{fig:original}) but with h$_{c}$ in the gas and small dust = 84 AU }
    \label{fig:hc84}
\end{figure}

\begin{figure}
    \centering
    \includegraphics[scale=.5]{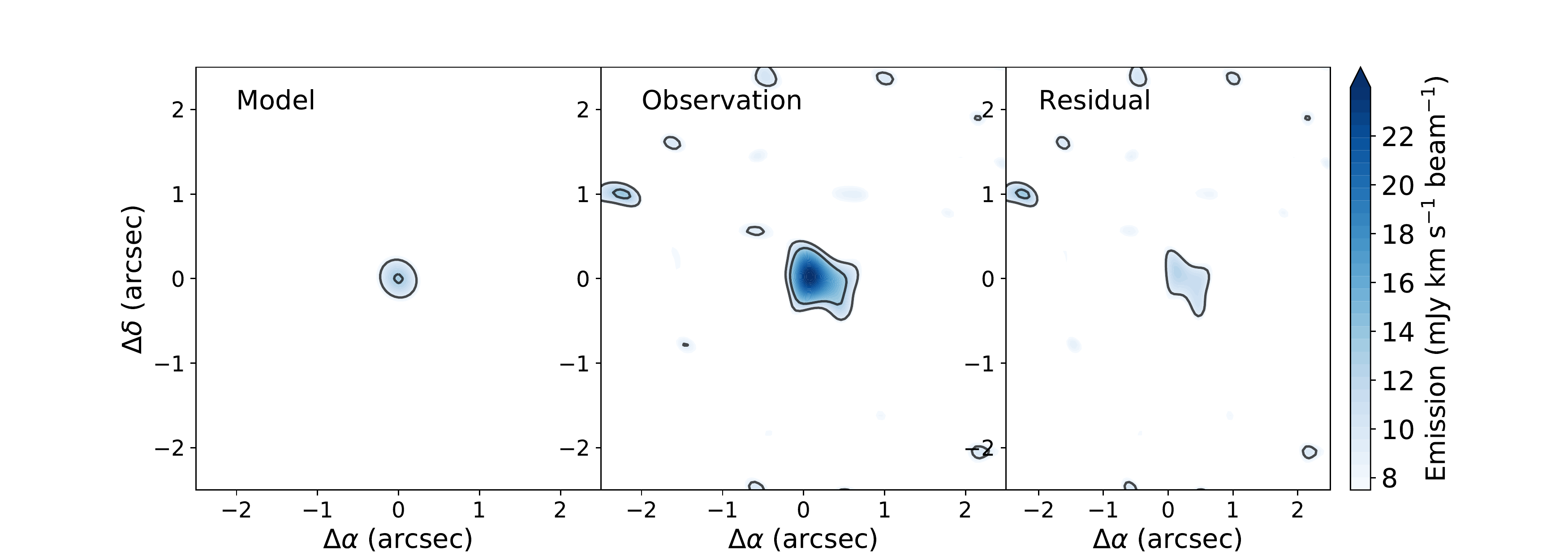}
    \caption{ALMA observation of $^{13}$C$^{18}$O $J$=3-2 line emission (middle panel) as compared to our final TW Hya modeled $^{13}$C$^{18}$O $J$=3-2 flux (left panel), and the residual (right panel). Contours are trace the location of 9 and 13 mJy flux values.}
    \label{13C18O}
\end{figure}



\newpage
\bibliography{references}{}
\bibliographystyle{aasjournal}



\end{document}